\documentclass[aps, prl, reprint, superscriptaddress, showpacs,preprintnumbers,amsmath,amssymb,twocolumn]{revtex4-2}

\usepackage{amssymb}
\usepackage{amsmath}
\usepackage{array}
\usepackage{tabularx}
\usepackage{bm}
\usepackage[dvips]{graphicx}
\usepackage{dcolumn}
\usepackage{longtable}
\usepackage{color}
\usepackage[version = 4]{mhchem}

\newcommand{\casb}{CaSb$_2$}
\newcommand{\Tc}{$T_{\mathrm{c}}$}
\newcommand{\Pmax}{$P_{\mathrm{max}}$}

\begin{document}
\title{Subtle Structural Anomaly under Compression in Line-Nodal CaSb$_2$}
\author{Hidemitsu~Takahashi}
\email{takahashi.hidemitsu.23r@st.kyoto-u.ac.jp}
\affiliation{Department of Physics, Graduate School of Science, 
Kyoto University, Kyoto 606-8502, Japan}

\author{Atsutoshi~Ikeda}
\affiliation{Department of Electronic Science and Engineering, Graduate School of Engineering, Kyoto University, Kyoto 615-8510, Japan}

\author{Shunsaku~Kitagawa}
\affiliation{Department of Physics, Graduate School of Science, 
Kyoto University, Kyoto 606-8502, Japan}

\author{Hirokazu~Kadobayashi}
\affiliation{Diffraction and Scattering Division, Center for Synchrotron Radiation, Japan Synchrotron Radiation Research Institute, Sayo, Hyogo 679-5198, Japan}

\author{Naohisa~Hirao}
\affiliation{Diffraction and Scattering Division, Center for Synchrotron Radiation, Japan Synchrotron Radiation Research Institute, Sayo, Hyogo 679-5198, Japan}

\author{Kenji~Ishida}
\affiliation{Department of Physics, Graduate School of Science, 
Kyoto University, Kyoto 606-8502, Japan}

\author{Tsuyoshi~Imazu}
\affiliation{Department of Mathematics and Physics, Hirosaki University, Hirosaki, Aomori 036-8561, Japan}

\author{Yoshiteru~Maeno}
\affiliation{Toyota Riken-Kyoto University Research Center (TRiKUC), Kyoto 606-8501, Japan}

\begin{abstract}
    We report X-ray diffraction patterns and calculated electronic band structures of the Dirac line-nodal material \casb ~under pressure.
    Its superconducting transition temperature ($T_{\mathrm{c}}=1.7$ K) increases under pressure and reaches a maximum at 3.4 K at around 3 GPa. 
    We observed subtle anomalies in lattice parameters accompanied by a jump in bulk modulus without any change in crystal symmetry at around 3 GPa.
    First-principles calculations revealed that the distorted lattice of Sb(1) site deforms in the pressure range of 0-3 GPa.
    Those results suggest the existence of a first-order structural transition and arouse expectations for unusual phononic properties affecting the superconducting state.
    The calculated pressure dependence of the electronic density of states (DOS) confirms that it is not the change in the DOS that governs the variations in \Tc. 
        
\end{abstract}

\maketitle

\section{Introduction}
Topological line-nodal material has become the platform of active research in the last decade.
In such a system, topologically protected band crossings exist as a line in the Brillouin zone (BZ),
resulting in the relativistic quasiparticles and novel quantum phenomena \cite{PhysRevB.93.035138,hirayamaTopologicalDiracNodal2017,PhysRevB.95.075138,PhysRevB.97.161113,PhysRevResearch.2.043311}.
Generally, the protection of nodal lines requires certain crystal symmetries including mirror and nonsymmorphic symmetries \cite{fangTopologicalNodalLine2016}, 
and thus the crystal structure plays crucial roles in line-nodal materials.

\casb ~has a nonsymmorphic crystalline symmetry and is a candidate for the Dirac line-nodal material \cite{funadaSpinOrbitCoupling2019}.
In the Dirac line-nodal material, the band crossing points are the Dirac points.
The crystal structure of \casb ~belongs to a monoclinic space group of $P2_1/m$ (No.11, $C_{2h}^2$) with screw symmetry.
There exist two nonequivalent Sb sites, Sb(1) and Sb(2), as shown in Figs. \ref{fig:structure}(a) and \ref{fig:structure}(b).
Sb(1) atoms form zigzag chains along the $b$ axis and the chains form distorted square lattices.
As pointed out in the literature \cite{PhysRevB.105.184504}, diantimonides $M$Sb$_2$ ($M=$~rare earth, alkaline earth) including \casb ~can be considered as 
a variation of a family of line-nodal materials with a square net such as ZrSiS \cite{schoopDiracConeProtected2016}, LaSbTe \cite{PhysRevB.103.125131}, and GdSbTe \cite{hosenDiscoveryTopologicalNodalline2018}.
The nodal lines in \casb, protected by the nonsymmorphic crystalline symmetry, are robust against spin-orbit coupling (SOC).
First-principles-calculation study reported that the Fermi surfaces mainly originate from Sb(1) and Sb(2) \cite{funadaSpinOrbitCoupling2019}.
Sb(1) is dominant in the nearly two-dimensional electronic Fermi surfaces forming the nodal lines, while Sb(2) forms three-dimensional hole surface around the $\Gamma$ point.
Quantum oscillations and angle-resolved photoemission study reported the electronic band structure consistent with the first-principles calculations \cite{ikedaQuasitwodimensionalFermiSurface2022,chuangFermiologyTopologicalLinenodal2022}.
Therefore, the configuration of the Sb(1) atoms would be especially important for the topological properties of \casb.

Another feature of \casb ~is its superconductivity below $T_{\mathrm{c}}=1.7$ K \cite{ikedaSuperconductivityNonsymmorphicLinenodal2020}.
Most of the experimental results support conventional full-gap superconductivity at ambient pressure \cite{takahashiSWaveSuperconductivityDirac2021,PhysRevB.106.214521,chuangFermiologyTopologicalLinenodal2022,ikedaSuppressionSuperconductingTransition2024}.
However, the superconducting (SC) properties under pressure are not simple.
As illustrated in Fig.~\ref{fig:structure}(c), \Tc ~shows a peak structure under hydrostatic pressure with a maximum value of 3.4 K at 3.1 GPa \cite{kitagawaPeakSuperconductingTransition2021,PhysRevB.109.L100501}.
The decrease in \Tc ~by chemical substitution attributable to the effect of the negative pressrue was also observed \cite{ikedaSuppressionSuperconductingTransition2024}.
In a conventional weak-coupling superconductor, \Tc ~monotonically decreases by pressure because the electronic density of states (DOS) at the Fermi energy $N(E_{\mathrm{F}})$ decreases due to the
increasing bandwidth.
In reality, a previous work of $^{123}$Sb-nuclear quadrupole resonance (NQR) measurements under pressure up to 2.08 GPa revealed that $N(E_{\mathrm{F}})$ is insensitive to pressure in \casb 
~and is not the origin of the increase in \Tc ~\cite{PhysRevB.109.L100501}.
According to the Bardeen-Cooper-Schrieffer (BCS) theory, the remaining possibility of the key factor for the increasing \Tc ~is the Debye frequency $\omega_{\mathrm{D}}$
or the effective electronic interaction, both being related to the crystal lattice.
Although the unusual broadening in the $^{123}$Sb-NQR spectrum suggests the unique features of the compression of \casb, the origin of the increasing \Tc ~is still elusive.
Moreover, \Tc ~begins to decrease by applying pressure greater than 3.1 GPa.
The nonmonotonic behavior of \Tc ~against pressure with a peak at $P_{\mathrm{max}}=3.1$ GPa is indicative of unconventional character of the SC state in \casb ~and 
considered to be important for understanding the SC phenomena in the topological line-nodal materials.

To identify the origin of such unusual pressure dependence in \casb, the information on the crystal structure under pressure is indispensable.
It is also crucial for the topological properties of \casb ~because the nodal lines are protected by the crystalline symmetry.
However, experimental results of the lattice properties under pressure have not been reported so far.
Therefore, we carried out high-resolution synchrotron X-ray diffraction (XRD) measurements under pressure to investigate possible
structural anomaly with a limited amount of sample.

In this paper, we report the results of XRD measurements under pressure along with the first-principles calculations.
Although no symmetry change was observed, subtle anomalies at 3 GPa, where \Tc ~reaches the maximum, were observed in the pressrue dependence of the lattice parameters.
The crystal lattice exhibits softening at \Pmax, characterizing the unique feature of the compression of \casb.
The first-principles calculations revealed that the anisotropy of shrinkage changes at \Pmax, which possibly affects the phononic properties and the SC state.
The calculated pressure dependence of DOS at $E_{\mathrm{F}}$ confirmed that the peak in \Tc ~is completely unattributable to DOS. 
The pressure evolution of nodal lines was also investigated. The existence of nodal lines is robust against pressure because of the absence of the symmetry change.
The reconnection of nodal lines was observed in the high pressure region.

\section{Experiment and Calculation}

\begin{figure}[tbp]
    \centering
    \includegraphics[width=8.6cm]{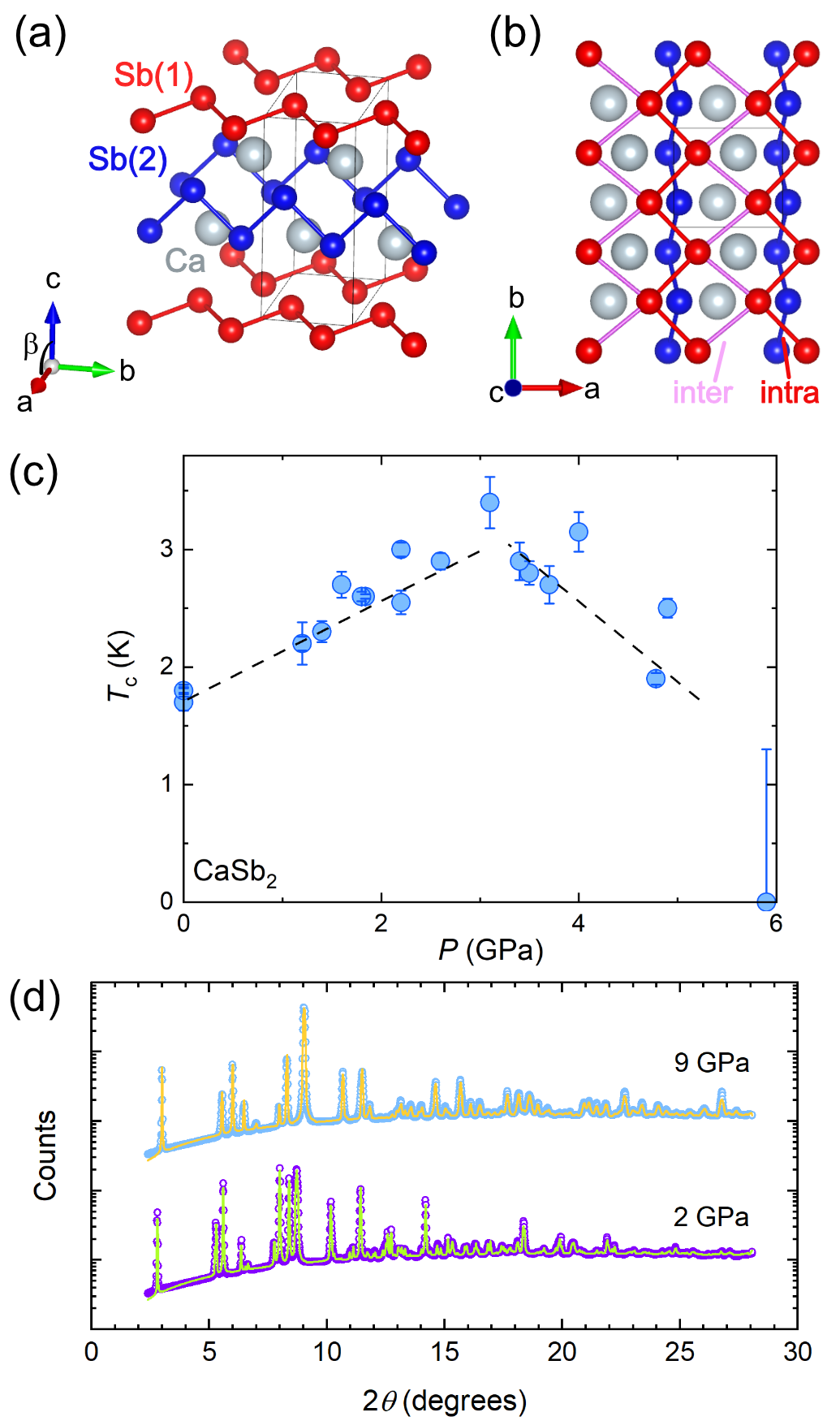}
    \caption{(Color online) 
    (a) Crystal structure of \casb ~drawn by VESTA \cite{Momma:db5098}. (b) Distorted square lattice of Sb(1) atoms. Intrachain (interchain) bondings of Sb(1) are colored in red (pink).
    (c) Pressure dependence of \Tc ~adapted from Ref.~\cite{kitagawaPeakSuperconductingTransition2021}.
    (d) XRD patterns at various pressures. Circles represent the data points, and the curves show the fit.}
    \label{fig:structure}
\end{figure}

Single crystals of \casb ~were prepared by the self-flux method and powdered in argon atmosphere.
Ca (Sigma-Aldrich, 99.99\%) and Sb (Rare Metallic, 99.9999\%) with a molar ratio of 1:4 were heated in a box furnace (Denken, KDF 80S) 
using a temperature profile similar to that reported in Ref. 12 but with the final temperature of 600 $^{\circ}$C.
XRD measurement at ambient pressure was performed at room temperature with a commercial diffractometer (Bruker AXS, D8 Advance) using Cu-K$\alpha$ radiation.
Synchrotron XRD measurements at room temperature under pressure were carried out at BL10XU beamline of SPring-8, Japan \cite{hiraoNewDevelopmentsHighpressure2020}. 
The energy of the incident beam is 30 keV. 
Hydrostatic pressure was applied to the sample using a helium-gas-driven membrane-type diamond anvil cell (DAC) with a culet diameter of 500 $\mu$m from 0.29 to 9.71 GPa.
Helium gas was used as a pressure medium.
The powder sample was loaded to a gasket made of SUS301 in air and the diameter of the sample was 260 $\mu$m.
Pressure inside the cell was monitored by fluorescence spectra of single crystalline ruby balls \cite{https://doi.org/10.1029/JB091iB05p04673}.
The pressure values before and after the XRD measurements are averaged to determine the applied pressure. 
Two-dimensional diffraction images were obtained using an imaging plate detector and integrated along the angular direction 
to make them conventional one-dimensional profiles shown in Fig.~\ref{fig:structure}(d) \cite{setoDevelopmentSoftwareSuite2010}. Lattice parameters were fitted using the TOPAS software by the Rietveld analysis \cite{Rietveld:a07067} for the data at ambient pressure 
and by the whole-powder-pattern decomposition using the Pawley method \cite{Pawley:a20546} for the data under pressure.

First-principles electronic structure calculations based on the density functional theory (DFT) were performed using the QUANTUM ESPRESSO package \cite{giannozziQUANTUMESPRESSOModular2009,giannozziAdvancedCapabilitiesMaterials2017}.
We adopted fully relativistic projector augmented wave pseudopotentials with Perdew-Burke-Ernzerhof exchange-correlation functional \cite{PhysRevLett.77.3865} from pslibrary 1.0.0 \cite{dalcorsoPseudopotentialsPeriodicTable2014}.
SOC was included self-consistently.
The cutoff energy of the plane-wave basis was set to be 45 Ry.
Experimentally obtained lattice parameters were used, and the internal degrees of freedom of the atomic positions were relaxed until the forces 
on each atom became smaller than $10^{-4}$ Ry/$a_0$, where $a_0$ is the Bohr radius.
Monkhorst-Pack $k$ mesh of $20\times22\times12$ was employed for charge-density calculations.
The DOS and Fermi surfaces were calculated with further interpolated $k$-mesh of $40\times44\times24$.
We constructed maximally localized Wannier functions using Wannier90 \cite{mostofiUpdatedVersionWannier902014} from $p_x$, $p_y$, and $p_z$ orbitals of the Sb(1) and Sb(2) atoms.
The location of the nodal lines was calculated by WannierTools package \cite{wuWannierToolsOpensourceSoftware2018}.

\section{Results and Discussion}

\begin{figure}[tbp]
    \centering
    \includegraphics[width=8.6cm]{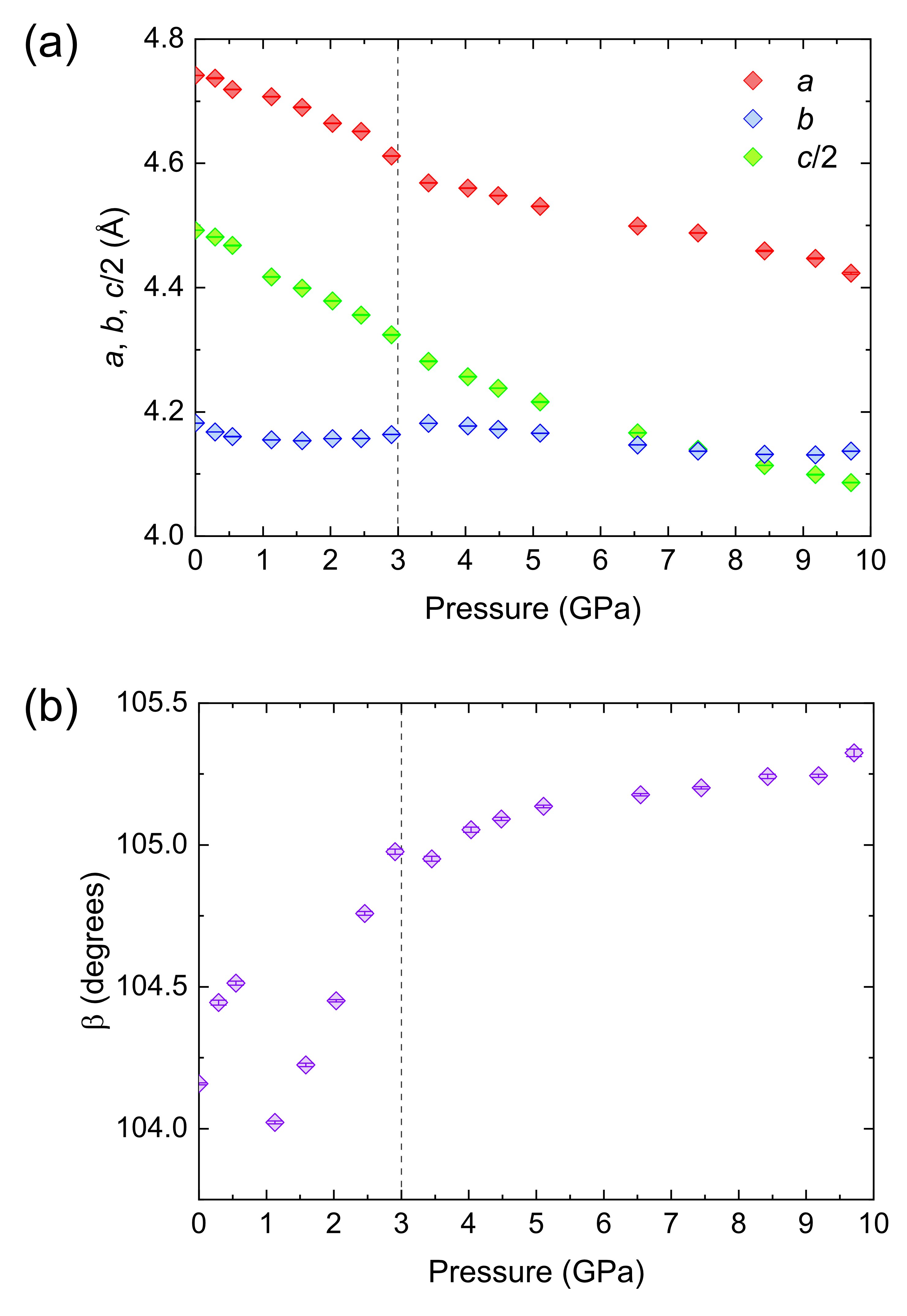}
    \caption{(Color online) 
    Pressure dependence of the lattice parameters (a) $a$, $b$, and $c/2$ and
    (b) the angle $\beta$ shown in Fig.~\ref{fig:structure}(a). Broken line indicates 3 GPa ($\sim$ \Pmax).}
    \label{fig:xray}
\end{figure}

\begin{figure}[tbp]
    \centering
    \includegraphics[width=8.6cm]{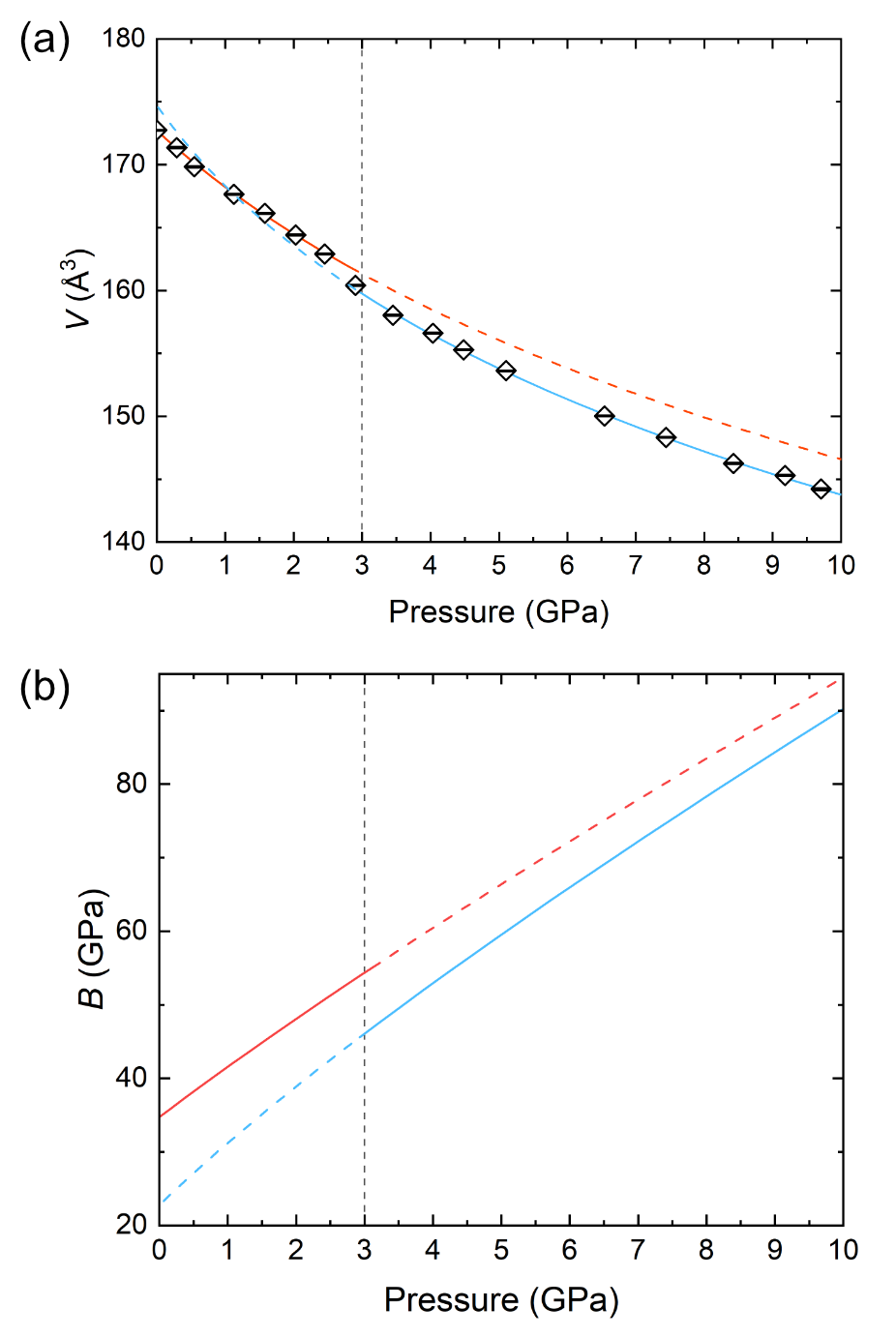}
    \caption{(Color online) 
    (a) Pressure dependence of the unit cell volume $V$ along with the fitting by Eq.~\eqref{Murgnahan} for low $P$ (red) and high $P$ (blue) regions presented in Table~\ref{Murgnahanfit}.
    (b) Calculated bulk modulus using Eq.~\eqref{bulkmodulus} for low $P$ (red) and high $P$ (blue) regions.}
    \label{fig:Murgnahan}
\end{figure}

\subsection{XRD}
Figure~\ref{fig:structure}(d) shows the representative XRD profiles measured below and above \Pmax.
The Bragg peaks were well indexed by the space group $P2_1/m$ in the entire pressure region with almost constant goodness-of-fit indicator (GOF$\sim$3).
Based on the result, hydrostatic pressure does not induce symmetry change in the crystal lattice of \casb ~up to 9.71 GPa.
The pressure dependence of the extracted lattice parameters is shown in Fig.~\ref{fig:xray}.
The lengths of the $a$ and the $c$ axes monotonically decrease by applying pressure while that of the $b$ axis is almost independent of pressure, reflecting the important role of 
the Sb(1) chains in the lattice structure.
Notably, all the lattice parameters exhibit subtle anomalies at around \Pmax, suggesting the existence of a transition without symmetry change.
The angle $\beta$ between the $a$ and $c$ axes shows the monotonic increase in most part and its slope against pressure changes at around \Pmax, coinciding with the anomalies in 
the lengths of crystal axes.
The pressure dependence of the unit cell volume $V$ also shows a jump-like anomaly at around \Pmax, ~as shown in Fig.~\ref{fig:Murgnahan}(a).
It is natural to consider that these anomalies at \Pmax~in the lattice are related to the nonmonotonic change of \Tc.
Here, we comment on the nonmonotonic behavior of $\beta$ between 0 and 1.1 GPa. 
We believe that it is not due to measurement error, but its origin is not clear.
Whatever it is, the rate of change of $\beta$ is small, so its effect on the crystal lattice
would not be significant. In fact, the pressure dependences of the interplanar distances of the unit cell $a\sin\beta$ and $c\sin\beta$ are almost the same as those of $a$ and $c$, respectively.
We also see little anomaly at 1.1 GPa in the \(P-V\) curve.

In order to extract the qualitative difference expected from the anomaly, we separately fitted the pressure dependence of $V$ below and above \Pmax ~to 
the third order Birch-Murgnahan equation of state used for isothermal process of solids \cite{PhysRev.71.809}
\begin{multline}
    P(V)=\frac{3B_0}{2}\left[\left(\frac{V}{V_0}\right)^{-\frac{7}{3}}-\left(\frac{V}{V_0}\right)^{-\frac{5}{3}}\right]\\
    \times\left\{1+\frac{3}{4}(B_0'-4)\left[\left(\frac{V}{V_0}\right)^{-\frac{2}{3}}-1\right]\right\},\label{Murgnahan}
\end{multline}
where $B_0$ is bulk modulus at 0 GPa, $B_0'$ is the derivative of $B_0$ with respect to pressure at 0 GPa, and $V_0$ is the volume at 0 GPa.
The data point of 2.91 GPa is not used for the fittings, because it is right in the middle of the anomaly.
The obtained curves and fitting parameters are shown in Fig.~\ref{fig:Murgnahan}(a) and Table~\ref{Murgnahanfit}, respectively. 
From the definition of the bulk modulus
\begin{equation}
    B = -V\frac{\partial P}{\partial V}
\end{equation}
and Eq.~\eqref{Murgnahan}, we obtain the expression of $B$ as
\begin{multline}
    B=\frac{3B_0}{2}\left[\frac{7}{3}\left(\frac{V}{V_0}\right)^{-\frac{7}{3}}-\frac{5}{3}\left(\frac{V}{V_0}\right)^{-\frac{5}{3}}\right.\\
   \left. +(B_0'-4)\left\{\frac{9}{4}\left(\frac{V}{V_0}\right)^{-3}-\frac{7}{2}\left(\frac{V}{V_0}\right)^{-\frac{7}{3}}+\frac{5}{4}\left(\frac{V}{V_0}\right)^{-\frac{5}{3}}\right\}\right]. \label{bulkmodulus}
\end{multline}
Using Eq.~\eqref{Murgnahan},~\eqref{bulkmodulus} and parameters in Table~\ref{Murgnahanfit}, we derived the pressure dependence of the bulk modulus as presented in Fig.~\ref{fig:Murgnahan}(b).
$B$ monotonically increases up to \Pmax ~and abruptly drops at \Pmax, showing monotonic increase again with further pressure but with a larger slope.
The discontinuity in $B$ more clearly gives an indication of the existence of a transition in the lattice, although the change in the lattice parameters is subtle.
This discontinuity in \(B\) leads to a jump in the \(P-V\) curve as seen in the result of the fitting.
If there is a structural transition, 
it would be first order since \(V\) exhibits a discontinuous jump with no change in the crystal symmetry~\cite{Christy:ab0333}.
As $B$ is related to the hardness of a material, the sudden drop of $B$ indicates the softening of the lattice at \Pmax.
In terms of $B$, the high-pressure phase is always softer than the low-pressure phase in the measured pressure range.
This behavior is unusual because crystal lattices tend to get harder by compression.
The increase in $B$ ~below \Pmax, which would result in the increase in $\omega_{\mathrm{D}}$, is consistent with the enhancement of \Tc ~below \Pmax.
However, whether the increase in \Tc ~mainly comes from $\omega_{\mathrm{D}}$ is uncertain because quantitative discussion requires information about elastic constants other than $B$
in monoclinic systems.
Furthermore, the decrease in \Tc ~above \Pmax ~requires a factor other than $\omega_{\mathrm{D}}$ governinig \Tc.

We point out that similar behavior of the $c$ axis is observed in the half-collapse transition in an iron-based superconductor CaKFe$_4$As$_4$ \cite{PhysRevB.96.140501}.
In such a transition, the length of the $c$ axis suddenly decreases due to the development of As-As bond across a Ca layer along the $c$ axis accompanied by the 
abrupt enlargement of the tetragonal plane.
This transition also preserves the crystal symmetry.
In CaKFe$_4$As$_4$, the bulk SC state is suddenly and discontinuously surpressed after the half-collapse transition.
In contrast, in our case, the absolute value of the slope of \Tc ~is almost the same below and above \Pmax~\cite{kitagawaPeakSuperconductingTransition2021}. 

The pressure evolution of the atomic positions might play an important role, but we could not determine them from the experimental data because the intensity ratio of 
the XRD peaks taken with a pressure cell was different from those taken at ambient pressure.
Given that the single crystal of \casb ~grows in a plate-like shape, this difference is explained by the selective orientation of the powdered sample within the DAC.
For further insight, we calculated the atomic coordinates of \casb ~by DFT using the lattice parameters obtained from XRD measurements.

\subsection{Atomic Positions}
\begin{figure}[tbp]
    \centering
        \includegraphics[width=8.6cm]{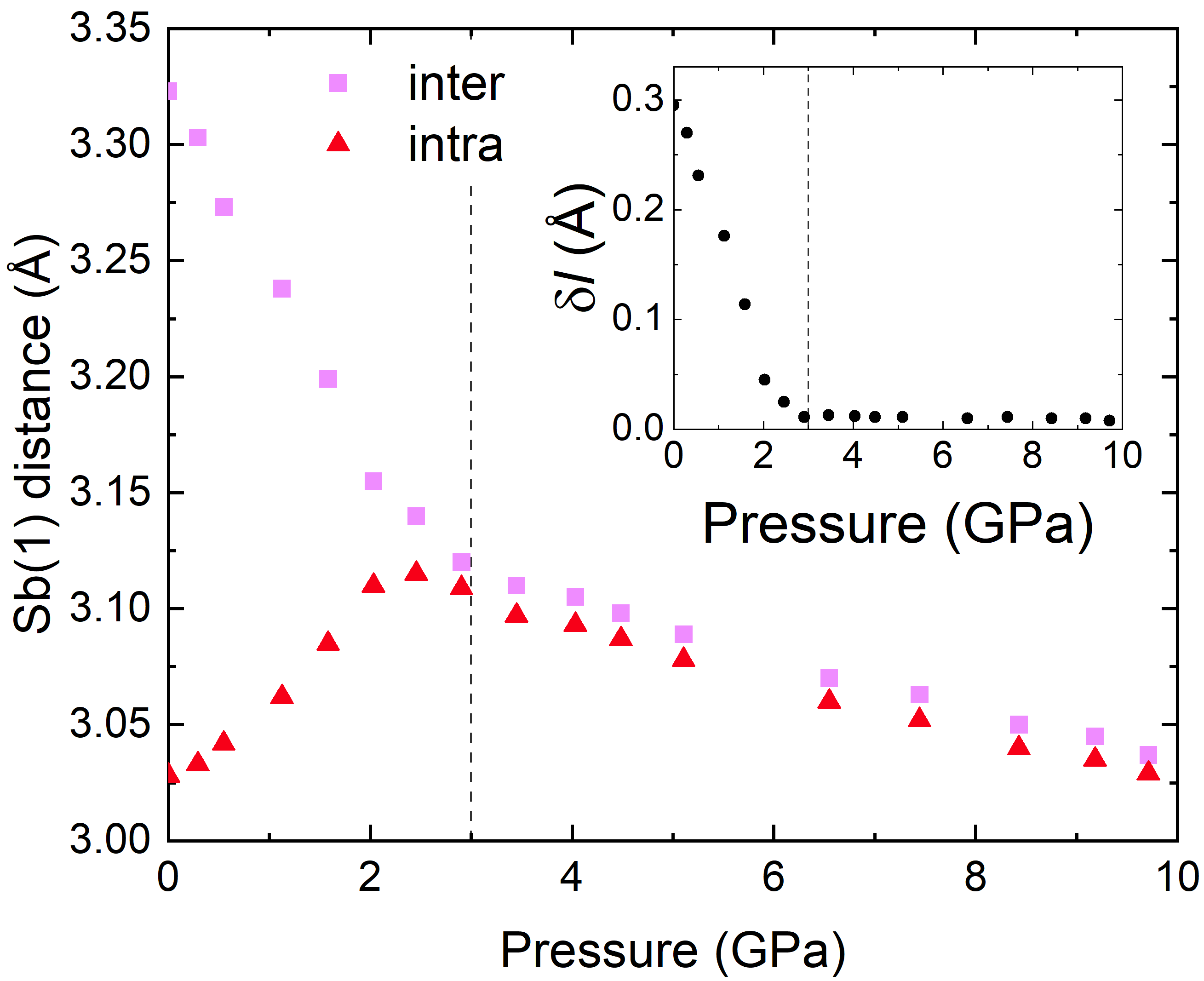}
        \caption{(Color online) 
        Pressure dependence of the intra- and inter-chain distances of Sb(1) atoms. Inset: pressure dependence of $\delta l=l_{\mathrm{inter}}-l_{\mathrm{intra}}$.  }
        \label{fig:deform}
\end{figure}

We used the experimentally obtained lattice parameters as inputs and relaxed the atomic degrees of freedom by DFT calculations.
In \casb, all the atoms occupy the Wyckoff position of $2e$.
The symmetry of the atomic positions is unchanged at any pressure and 
the pressure evolution of the crystal structure is revealed to be characterised by the deformation of the distorted square lattice formed by the Sb(1) chains shown in Fig.~\ref{fig:structure}(b).
Figure \ref{fig:deform} shows the pressure dependence of the distance between the Sb(1) sites within a zigzag chain ($l_{\mathrm{intra}}$) and those across the chain ($l_{\mathrm{inter}}$).
The difference between these distances $\delta l=l_{\mathrm{inter}}-l_{\mathrm{intra}}$ is also shown in the inset of Fig.~\ref{fig:deform}.
At ambient pressure, $\delta l$ is finite and the Sb(1) lattice is far from rhombus.
As pressure is applied, the intra-chain distance increases while the inter-chain distance decreases.
$\delta l$ monotonically decreases and reaches almost zero at 3 GPa, although the lattice is still distorted in the out-of-plane direction.
Above 3 GPa the lattice shrinks without the deformation of the Sb(1) lattice, preserving the original symmetry of the crystal.
Therefore, below and above \Pmax, the response of the crystal lattice against pressure is different.
This behavior suggests that the one-dimensional Sb(1) chains evolve into the two-dimensional network at \Pmax.
Such evolution is accompanied by a rather continuous change in the bond lengths, which confirms that the case is different from CaKFe$_4$As$_4$.
Considering that the $c$ axis shrinks the most and has almost identical pressure dependence with $V$, 
the softening above \Pmax ~would be explained by the promotion of the shrinkage along $c$ due to the saturation of the in-plane distortion.
In our previous work, we pointed out that phonon anomaly may play a role in the enhancement of \Tc ~in this system \cite{PhysRevB.109.L100501}.
The deformation of the lattice with the softening would affect the phonon properties, and thus, our result is in line with the above scenario.

\begingroup
\renewcommand{\arraystretch}{1.8}
\begin{table}
    \centering
    \caption{The volume $V_0$ at 0 GPa, bulk modulus $B_0$ at 0 GPa, and its derivative $B_0'$ at 0 GPa of \casb ~obtained by fits with the third order Birch-Murgnahan equation of state
    for low $P$ (0-2.46 GPa) and high $P$ (3.45-9.71 GPa) regions.}
    \label{Murgnahanfit}
    \setlength{\tabcolsep}{6pt}
    \begin{tabularx}{7.0cm}{lccc}\hline
         & $V_0$ (\AA$^3$)& $B_0$ (GPa)& $B_0'$\\ \hline
         low $P$ & 172.67$\pm$0.18 & 34.7$\pm$3.1 & 7.0$\pm$3.0\\
         high $P$ & 174.7$\pm$4.0 & 23$\pm$11 & 9.1$\pm$4.0\\ \hline
    \end{tabularx}
\end{table}
\endgroup

\subsection{Electronic Structure}
\begin{figure}[tbp]
    \centering
    \includegraphics[width=8.6cm]{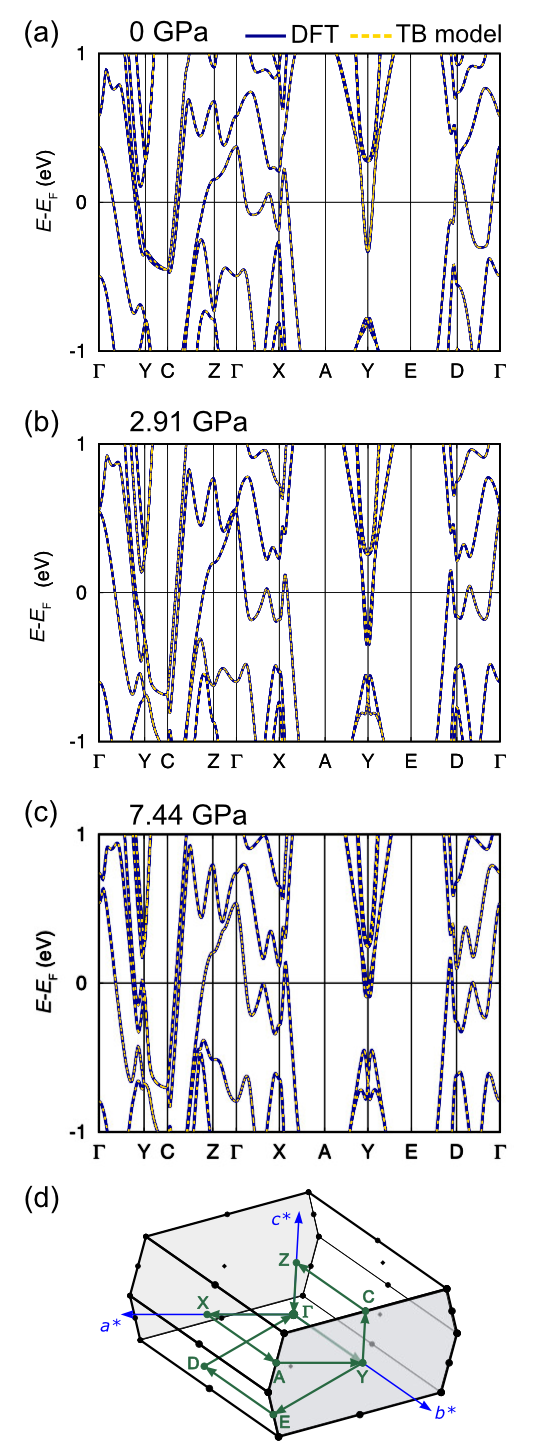}
    \caption{(Color online) 
    Electronic band structure calculated by DFT (dark blue) and TB model (yellow) at (a) 0 GPa, (b) 2.91 GPa, and (c) 7.44 GPa.
    (d) BZ of \casb. The green arrows indicate the $k$-path used in the band structure calculations. The zone boundaries corresponding to Fig.~\ref{fig:node} are colored in gray. }
    \label{fig:calc}
\end{figure}

We calculated the electronic structure and constructed the maximally localized Wannier functions based tight-binding (TB) model to examine the effect of compression on the electronic properties.
The band structures obtained from DFT calculations (dark blue) at 0, 2.91, and 7.44 GPa are shown in Figs.~\ref{fig:calc}(a), \ref{fig:calc}(b), and \ref{fig:calc}(c), respectively, where
the path in the reciprocal space is shown in Fig. \ref{fig:calc}(d). In addition, the band structures of TB models are presented (yellow).
Our TB models show excellent agreement with DFT results at all the pressures. 
Electronic bands have four-fold degeneracy at high symmetry points Y, C, A, and E forming the Dirac points, and those Dirac points are connected by nodal lines running within the 
BZ boundaries shown in Fig.~\ref{fig:calc}(d).
Applying pressure increases the band width by enhancing the hybridization of the electronic orbitals.
The energy dispersion along Y-A-X-$\Gamma$ and Y-E-D-$\Gamma$ become almost the same under pressure, indicating the increased symmetry around the $b^*$ axis.

\begin{figure}[tbp]
    \centering
    \includegraphics[width=8.6cm]{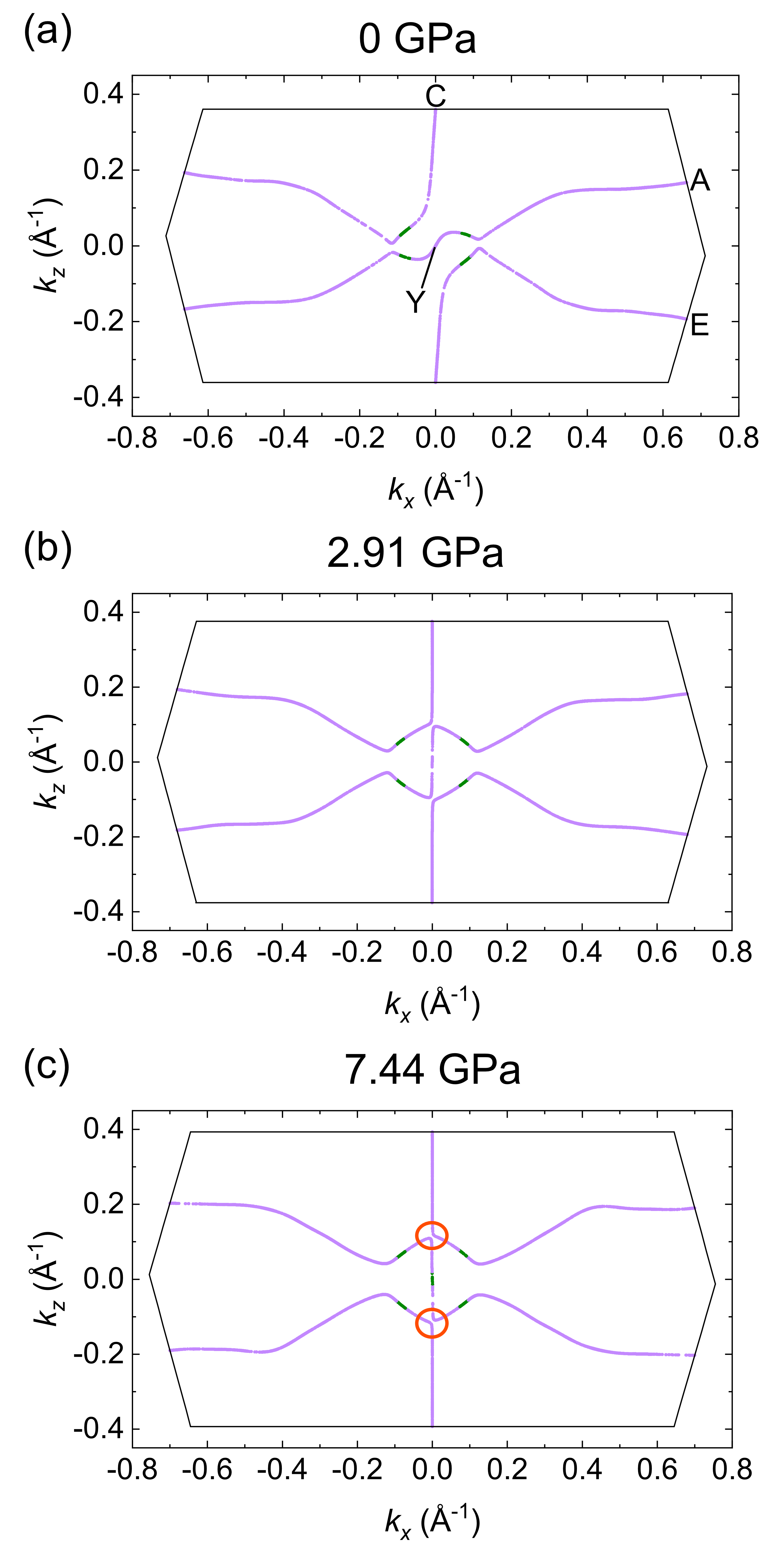}
    \caption{(Color online) 
    The location of the nodal lines at (a) 0 GPa, (b) 2.91 GPa, and (c) 7.44 GPa. The small green sections indicates the nodal points near $E_{\mathrm{F}}$.
    The orange circles indicate the region where the reconnection of the nodal lines occurs.
}
    \label{fig:node}
\end{figure}

Since Sb(1) shows characteristic response to pressure and is important for the nodal lines, we investigated the pressure evolution of the nodal lines using the TB models.
Figures \ref{fig:node}(a)-(c) show the calculated locations of the nodal lines at each pressure. 
Nodal lines have strong energy dispersion and cross the Fermi surface around the Y point.
The small regions included in $E_{\mathrm{F}}\pm0.1$ eV are colored in green.
Due to the preserved crystalline symmetry protecting the Dirac line nodes in the entire pressure region, the nodal lines are robust against pressure.
At 2.91 GPa, nodal lines align more symmetrically than at ambient pressure. This change is corresponding to the band structures under pressure.
By further compression to 7.44 GPa, the bands at the Y point get closer to $E_{\mathrm{F}}$.
At first glance, the contribution from the nodal lines to physical properties seems to become stronger in higher pressure region.
However, this is not the case because the density of nodal points around the Y point decreases due to the band expansion and 
the total number of Dirac points included in $E_{\mathrm{F}}\pm0.1$ eV does not increase.
Apart from that, the reconnection of the nodal lines was observed around the region marked by orange circles in Fig.~\ref{fig:node}(c).
This reconnection would be called a topological transition because the nodal lines are boundaries of the topological invariant defined by mirror operator \cite{funadaSpinOrbitCoupling2019}
and the reconnection requires the change in the topological invariant.
At a certain pressure in the process of reconnection, eight-fold degenerate points emerge.
The effect of these characteristic pressure evolution of the nodal lines on the properties of \casb ~should be attractive as future work.


\begin{figure}
    \centering
    \includegraphics[width=8.6cm]{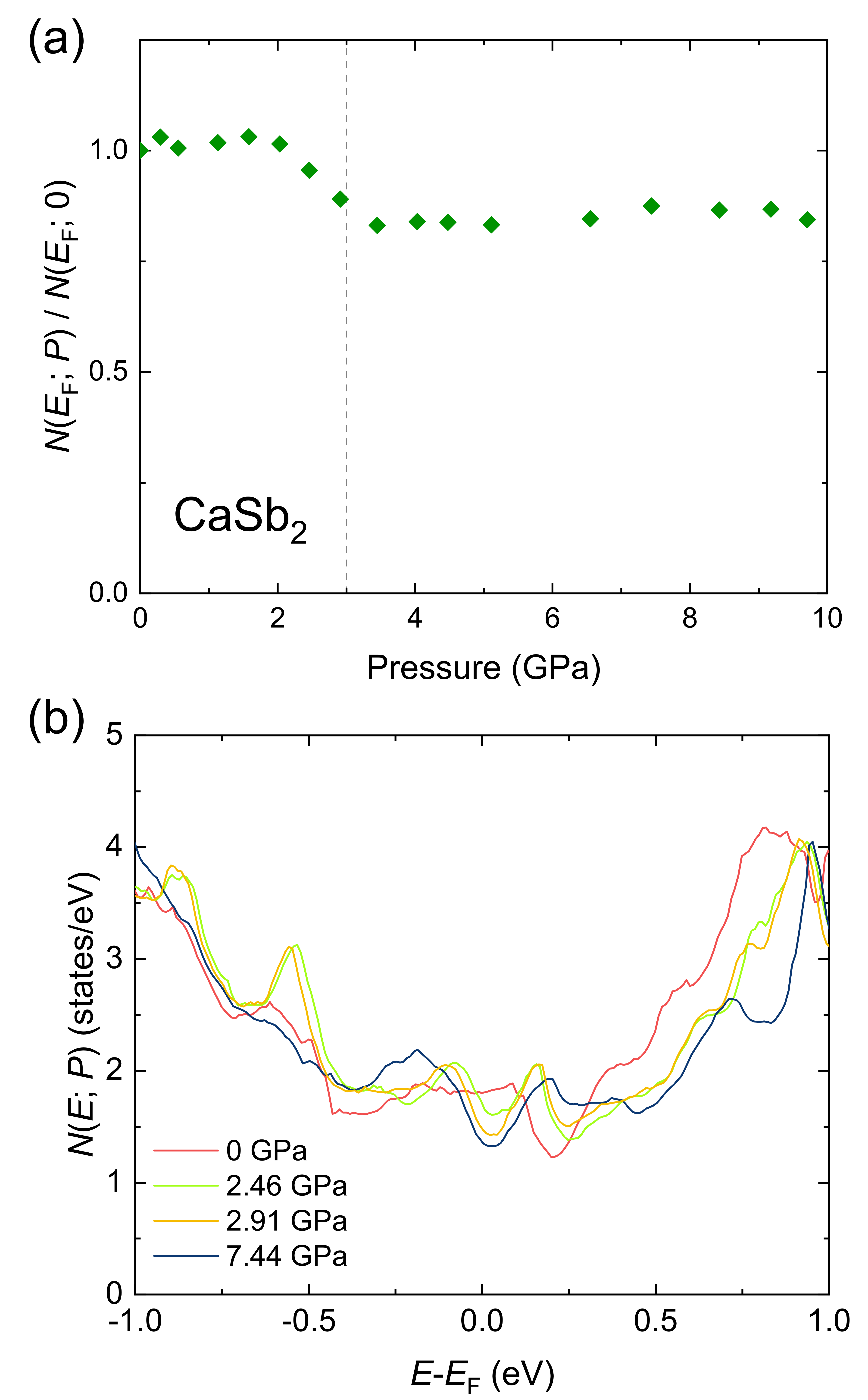}
    \caption{(Color online)
    (a) The DOS at $E_{\mathrm{F}}$ obtained by first-principles calculation normalized by the value at ambient pressure.
    (b) Energy dependence of DOS at 0, 2.46, 2.91, and 7.44 GPa.}
    \label{fig:dos}
\end{figure}

We calculated the electronic DOS at the Fermi energy at each pressure $N(E_{\mathrm{F}}; P)$.
The pressure dependence of $N(E_{\mathrm{F}}; P)$ normalized by the value at ambient pressure is shown in Fig.~\ref{fig:dos}(a).
$N(E_{\mathrm{F}}; P)$ is insensitive to pressure up to 2.03 GPa, consistent with the previous report of NQR measurements \cite{PhysRevB.109.L100501}.
From 2.03 to 3.45 GPa, $N(E_{\mathrm{F}}; P)$ linearly decreases to about 85\% of the value at ambient pressure.
As shown in Fig.~\ref{fig:dos}(b), a dip in $N(E; P)$ grows at $E_{\mathrm{F}}$ in this pressure range.
Above 3.45 GPa, the almost pressure independent $N(E_{\mathrm{F}}; P)$ is restored.
Clearly, the calculated pressure dependence of $N(E_{\mathrm{F}}; P)$ does not match that of \Tc.
Not only the initial increase in \Tc ~but also the decrease above \Pmax ~is revealed to be undescribable by $N(E_{\mathrm{F}}; P)$.

As we have seen, despite the structural anomaly, the electronic structure does not undergo any drastic change that would affect \Tc~at \Pmax.
Since $\omega_{\mathrm{D}}$ does not explain the pressure dependence of \Tc ~above \Pmax ~either, 
this seems to support the scenario that the electronic interaction mediated by phonons plays an important role for determining \Tc.
One system in which phonon modes associated with structural instability have a significant effect on electron-phonon coupling is a strong-coupling \(s\)-wave superconductor CaAlSi~\cite{PhysRevB.68.014512}.
The systematic study of solid solution Ca\(_{1-x}\)Sr\(_x\)AlSi revealed that only CaAlSi shows distortion from AlB\(_2\)-type structure and that \Tc~rapidly grows around \(x=0\)~\cite{PhysRevResearch.3.033192}.
Moreover, SrAlSi is a weak-coupling superconductor in contrast to CaAlSi~\cite{PhysRevB.68.014512}. 
The variation of the SC properties across the solid solution is considered to be caused by the enhanced electron-phonon coupling due to a phonon-softening related to the structural instability of CaAlSi.
Although the increase in \Tc~is not steep particularly around \Pmax~in \casb, the similar growth toward a strong-coupling SC state because of the structural anomaly might be observed.
Further research efforts to understand the nature of the SC state under pressure are desired.

\section{Conclusion}

In conclusion, we performed synchrotron XRD measurements under pressure and first-principles calculations to investigate the compression behavior and its effect on the 
electronic properties of \casb.
The XRD peaks were well fitted by the space group of $P2_1/m$ at all pressures and the subtle anomalies at \Pmax ~were observed in the pressure dependence of the lattice parameters, 
implying the first-order structural transition without symmetry change.
This transition might cause the softening of the crystal.
First-principles calculations revealed that Sb(1) square lattice distortion characterizes the compression below \Pmax, suggesting the transition from one-dimensional Sb(1) chain
to two-dimensional network.
These features predict the unusual phononic properties that affect the superconductivity.
If the transition of the Sb(1) network is crucial, uniaxial pressure along the $a$ axis may result in the deformation of Sb(1) lattice efficiently and realize the maximum \Tc ~below 3 GPa.
Nodal lines are robust against pressure but nevertheless exhibit a topological transition at the pressure between 2.91 and 7.44 GPa.
DOS was confirmed not to explain the nonmonotonic pressure dependence of \Tc, consistent with our previous study.
Especially above \Pmax, the effective electronic interaction mediated by phonons seems to play a key role for the change in \Tc.
Unveiling the nature of the discovered structural anomaly would be important for the understanding of the SC state of \casb ~under pressure.

\section{Acknowledgments}
The authors thank J. Goryo for valuable discussions.
The synchrotron radiation experiments were performed at the BL10XU of SPring-8 with the approval of 
Japan Synchrotron Radiation Research Institute (JASRI) (Proposal No. 2022B1729).
This work was supported by Grant-in-Aid for Scientific Research
on Innovative Areas from MEXT,
JSPS KAKENHI (Nos. JP20H00130,
JP20KK0061, JP21K18600, JP22H04933, JP22H01168, JP23H01124, JP23K22439, JP23K25821, JP24KJ1382, JP24K17011, and JP24H01659), the Kyoto University
Foundation, Toyota Physical and Chemical Research Institute, and the Murata Science Foundation. 
This work was also supported by JST SPRING (Grant Number JPMJSP2110 and JPMJSP2152).

\bibliographystyle{apsrev4-2}  
\bibliography{mylist}

\begin{thebibliography}{36}%
\makeatletter
\providecommand \@ifxundefined [1]{%
 \@ifx{#1\undefined}
}%
\providecommand \@ifnum [1]{%
 \ifnum #1\expandafter \@firstoftwo
 \else \expandafter \@secondoftwo
 \fi
}%
\providecommand \@ifx [1]{%
 \ifx #1\expandafter \@firstoftwo
 \else \expandafter \@secondoftwo
 \fi
}%
\providecommand \natexlab [1]{#1}%
\providecommand \enquote  [1]{``#1''}%
\providecommand \bibnamefont  [1]{#1}%
\providecommand \bibfnamefont [1]{#1}%
\providecommand \citenamefont [1]{#1}%
\providecommand \href@noop [0]{\@secondoftwo}%
\providecommand \href [0]{\begingroup \@sanitize@url \@href}%
\providecommand \@href[1]{\@@startlink{#1}\@@href}%
\providecommand \@@href[1]{\endgroup#1\@@endlink}%
\providecommand \@sanitize@url [0]{\catcode `\\12\catcode `\$12\catcode `\&12\catcode `\#12\catcode `\^12\catcode `\_12\catcode `\%12\relax}%
\providecommand \@@startlink[1]{}%
\providecommand \@@endlink[0]{}%
\providecommand \url  [0]{\begingroup\@sanitize@url \@url }%
\providecommand \@url [1]{\endgroup\@href {#1}{\urlprefix }}%
\providecommand \urlprefix  [0]{URL }%
\providecommand \Eprint [0]{\href }%
\providecommand \doibase [0]{https://doi.org/}%
\providecommand \selectlanguage [0]{\@gobble}%
\providecommand \bibinfo  [0]{\@secondoftwo}%
\providecommand \bibfield  [0]{\@secondoftwo}%
\providecommand \translation [1]{[#1]}%
\providecommand \BibitemOpen [0]{}%
\providecommand \bibitemStop [0]{}%
\providecommand \bibitemNoStop [0]{.\EOS\space}%
\providecommand \EOS [0]{\spacefactor3000\relax}%
\providecommand \BibitemShut  [1]{\csname bibitem#1\endcsname}%
\let\auto@bib@innerbib\@empty
\bibitem [{\citenamefont {Huh}\ \emph {et~al.}(2016)\citenamefont {Huh}, \citenamefont {Moon},\ and\ \citenamefont {Kim}}]{PhysRevB.93.035138}%
  \BibitemOpen
  \bibfield  {author} {\bibinfo {author} {\bibfnamefont {Y.}~\bibnamefont {Huh}}, \bibinfo {author} {\bibfnamefont {E.-G.}\ \bibnamefont {Moon}},\ and\ \bibinfo {author} {\bibfnamefont {Y.~B.}\ \bibnamefont {Kim}},\ }\href {https://doi.org/10.1103/PhysRevB.93.035138} {\bibfield  {journal} {\bibinfo  {journal} {Phys. Rev. B}\ }\textbf {\bibinfo {volume} {93}},\ \bibinfo {pages} {035138} (\bibinfo {year} {2016})}\BibitemShut {NoStop}%
\bibitem [{\citenamefont {Hirayama}\ \emph {et~al.}(2017)\citenamefont {Hirayama}, \citenamefont {Okugawa}, \citenamefont {Miyake},\ and\ \citenamefont {Murakami}}]{hirayamaTopologicalDiracNodal2017}%
  \BibitemOpen
  \bibfield  {author} {\bibinfo {author} {\bibfnamefont {M.}~\bibnamefont {Hirayama}}, \bibinfo {author} {\bibfnamefont {R.}~\bibnamefont {Okugawa}}, \bibinfo {author} {\bibfnamefont {T.}~\bibnamefont {Miyake}},\ and\ \bibinfo {author} {\bibfnamefont {S.}~\bibnamefont {Murakami}},\ }\href {https://doi.org/10.1038/ncomms14022} {\bibfield  {journal} {\bibinfo  {journal} {Nat. Commun.}\ }\textbf {\bibinfo {volume} {8}},\ \bibinfo {pages} {14022} (\bibinfo {year} {2017})}\BibitemShut {NoStop}%
\bibitem [{\citenamefont {Ramamurthy}\ and\ \citenamefont {Hughes}(2017)}]{PhysRevB.95.075138}%
  \BibitemOpen
  \bibfield  {author} {\bibinfo {author} {\bibfnamefont {S.~T.}\ \bibnamefont {Ramamurthy}}\ and\ \bibinfo {author} {\bibfnamefont {T.~L.}\ \bibnamefont {Hughes}},\ }\href {https://doi.org/10.1103/PhysRevB.95.075138} {\bibfield  {journal} {\bibinfo  {journal} {Phys. Rev. B}\ }\textbf {\bibinfo {volume} {95}},\ \bibinfo {pages} {075138} (\bibinfo {year} {2017})}\BibitemShut {NoStop}%
\bibitem [{\citenamefont {Rui}\ \emph {et~al.}(2018)\citenamefont {Rui}, \citenamefont {Zhao},\ and\ \citenamefont {Schnyder}}]{PhysRevB.97.161113}%
  \BibitemOpen
  \bibfield  {author} {\bibinfo {author} {\bibfnamefont {W.~B.}\ \bibnamefont {Rui}}, \bibinfo {author} {\bibfnamefont {Y.~X.}\ \bibnamefont {Zhao}},\ and\ \bibinfo {author} {\bibfnamefont {A.~P.}\ \bibnamefont {Schnyder}},\ }\href {https://doi.org/10.1103/PhysRevB.97.161113} {\bibfield  {journal} {\bibinfo  {journal} {Phys. Rev. B}\ }\textbf {\bibinfo {volume} {97}},\ \bibinfo {pages} {161113} (\bibinfo {year} {2018})}\BibitemShut {NoStop}%
\bibitem [{\citenamefont {Matsushita}\ \emph {et~al.}(2020)\citenamefont {Matsushita}, \citenamefont {Fujimoto},\ and\ \citenamefont {Schnyder}}]{PhysRevResearch.2.043311}%
  \BibitemOpen
  \bibfield  {author} {\bibinfo {author} {\bibfnamefont {T.}~\bibnamefont {Matsushita}}, \bibinfo {author} {\bibfnamefont {S.}~\bibnamefont {Fujimoto}},\ and\ \bibinfo {author} {\bibfnamefont {A.~P.}\ \bibnamefont {Schnyder}},\ }\href {https://doi.org/10.1103/PhysRevResearch.2.043311} {\bibfield  {journal} {\bibinfo  {journal} {Phys. Rev. Res.}\ }\textbf {\bibinfo {volume} {2}},\ \bibinfo {pages} {043311} (\bibinfo {year} {2020})}\BibitemShut {NoStop}%
\bibitem [{\citenamefont {Fang}\ \emph {et~al.}(2016)\citenamefont {Fang}, \citenamefont {Weng}, \citenamefont {Dai},\ and\ \citenamefont {Fang}}]{fangTopologicalNodalLine2016}%
  \BibitemOpen
  \bibfield  {author} {\bibinfo {author} {\bibfnamefont {C.}~\bibnamefont {Fang}}, \bibinfo {author} {\bibfnamefont {H.}~\bibnamefont {Weng}}, \bibinfo {author} {\bibfnamefont {X.}~\bibnamefont {Dai}},\ and\ \bibinfo {author} {\bibfnamefont {Z.}~\bibnamefont {Fang}},\ }\href {https://doi.org/10.1088/1674-1056/25/11/117106} {\bibfield  {journal} {\bibinfo  {journal} {Chine. Phys. B}\ }\textbf {\bibinfo {volume} {25}},\ \bibinfo {pages} {117106} (\bibinfo {year} {2016})}\BibitemShut {NoStop}%
\bibitem [{\citenamefont {Funada}\ \emph {et~al.}(2019)\citenamefont {Funada}, \citenamefont {Yamakage}, \citenamefont {Yamashina},\ and\ \citenamefont {Kageyama}}]{funadaSpinOrbitCoupling2019}%
  \BibitemOpen
  \bibfield  {author} {\bibinfo {author} {\bibfnamefont {K.}~\bibnamefont {Funada}}, \bibinfo {author} {\bibfnamefont {A.}~\bibnamefont {Yamakage}}, \bibinfo {author} {\bibfnamefont {N.}~\bibnamefont {Yamashina}},\ and\ \bibinfo {author} {\bibfnamefont {H.}~\bibnamefont {Kageyama}},\ }\href {https://doi.org/10.7566/JPSJ.88.044711} {\bibfield  {journal} {\bibinfo  {journal} {J. Phys. Soc. Jpn.}\ }\textbf {\bibinfo {volume} {88}},\ \bibinfo {pages} {044711} (\bibinfo {year} {2019})}\BibitemShut {NoStop}%
\bibitem [{\citenamefont {Oudah}\ \emph {et~al.}(2022)\citenamefont {Oudah}, \citenamefont {Bannies}, \citenamefont {Bonn},\ and\ \citenamefont {Aronson}}]{PhysRevB.105.184504}%
  \BibitemOpen
  \bibfield  {author} {\bibinfo {author} {\bibfnamefont {M.}~\bibnamefont {Oudah}}, \bibinfo {author} {\bibfnamefont {J.}~\bibnamefont {Bannies}}, \bibinfo {author} {\bibfnamefont {D.~A.}\ \bibnamefont {Bonn}},\ and\ \bibinfo {author} {\bibfnamefont {M.~C.}\ \bibnamefont {Aronson}},\ }\href {https://doi.org/10.1103/PhysRevB.105.184504} {\bibfield  {journal} {\bibinfo  {journal} {Phys. Rev. B}\ }\textbf {\bibinfo {volume} {105}},\ \bibinfo {pages} {184504} (\bibinfo {year} {2022})}\BibitemShut {NoStop}%
\bibitem [{\citenamefont {Schoop}\ \emph {et~al.}(2016)\citenamefont {Schoop}, \citenamefont {Ali}, \citenamefont {Stra{\ss}er}, \citenamefont {Topp}, \citenamefont {Varykhalov}, \citenamefont {Marchenko}, \citenamefont {Duppel}, \citenamefont {Parkin}, \citenamefont {Lotsch},\ and\ \citenamefont {Ast}}]{schoopDiracConeProtected2016}%
  \BibitemOpen
  \bibfield  {author} {\bibinfo {author} {\bibfnamefont {L.~M.}\ \bibnamefont {Schoop}}, \bibinfo {author} {\bibfnamefont {M.~N.}\ \bibnamefont {Ali}}, \bibinfo {author} {\bibfnamefont {C.}~\bibnamefont {Stra{\ss}er}}, \bibinfo {author} {\bibfnamefont {A.}~\bibnamefont {Topp}}, \bibinfo {author} {\bibfnamefont {A.}~\bibnamefont {Varykhalov}}, \bibinfo {author} {\bibfnamefont {D.}~\bibnamefont {Marchenko}}, \bibinfo {author} {\bibfnamefont {V.}~\bibnamefont {Duppel}}, \bibinfo {author} {\bibfnamefont {S.~S.~P.}\ \bibnamefont {Parkin}}, \bibinfo {author} {\bibfnamefont {B.~V.}\ \bibnamefont {Lotsch}},\ and\ \bibinfo {author} {\bibfnamefont {C.~R.}\ \bibnamefont {Ast}},\ }\href {https://doi.org/10.1038/ncomms11696} {\bibfield  {journal} {\bibinfo  {journal} {Nat. Commun.}\ }\textbf {\bibinfo {volume} {7}},\ \bibinfo {pages} {11696} (\bibinfo {year} {2016})}\BibitemShut {NoStop}%
\bibitem [{\citenamefont {Wang}\ \emph {et~al.}(2021)\citenamefont {Wang}, \citenamefont {Qian}, \citenamefont {Yang}, \citenamefont {Chen}, \citenamefont {Li}, \citenamefont {Tan}, \citenamefont {Cai}, \citenamefont {Zhao}, \citenamefont {Gao}, \citenamefont {Feng}, \citenamefont {Kumar}, \citenamefont {Schwier}, \citenamefont {Zhao}, \citenamefont {Weng}, \citenamefont {Shi}, \citenamefont {Wang}, \citenamefont {Song}, \citenamefont {Huang}, \citenamefont {Shimada}, \citenamefont {Xu}, \citenamefont {Zhou},\ and\ \citenamefont {Liu}}]{PhysRevB.103.125131}%
  \BibitemOpen
  \bibfield  {author} {\bibinfo {author} {\bibfnamefont {Y.}~\bibnamefont {Wang}}, \bibinfo {author} {\bibfnamefont {Y.}~\bibnamefont {Qian}}, \bibinfo {author} {\bibfnamefont {M.}~\bibnamefont {Yang}}, \bibinfo {author} {\bibfnamefont {H.}~\bibnamefont {Chen}}, \bibinfo {author} {\bibfnamefont {C.}~\bibnamefont {Li}}, \bibinfo {author} {\bibfnamefont {Z.}~\bibnamefont {Tan}}, \bibinfo {author} {\bibfnamefont {Y.}~\bibnamefont {Cai}}, \bibinfo {author} {\bibfnamefont {W.}~\bibnamefont {Zhao}}, \bibinfo {author} {\bibfnamefont {S.}~\bibnamefont {Gao}}, \bibinfo {author} {\bibfnamefont {Y.}~\bibnamefont {Feng}}, \bibinfo {author} {\bibfnamefont {S.}~\bibnamefont {Kumar}}, \bibinfo {author} {\bibfnamefont {E.~F.}\ \bibnamefont {Schwier}}, \bibinfo {author} {\bibfnamefont {L.}~\bibnamefont {Zhao}}, \bibinfo {author} {\bibfnamefont {H.}~\bibnamefont {Weng}}, \bibinfo {author} {\bibfnamefont {Y.}~\bibnamefont {Shi}}, \bibinfo {author} {\bibfnamefont {G.}~\bibnamefont {Wang}}, \bibinfo {author} {\bibfnamefont {Y.}~\bibnamefont {Song}}, \bibinfo {author} {\bibfnamefont {Y.}~\bibnamefont {Huang}}, \bibinfo {author} {\bibfnamefont {K.}~\bibnamefont {Shimada}}, \bibinfo {author} {\bibfnamefont {Z.}~\bibnamefont {Xu}}, \bibinfo {author} {\bibfnamefont {X.~J.}\ \bibnamefont {Zhou}},\ and\ \bibinfo {author} {\bibfnamefont {G.}~\bibnamefont {Liu}},\ }\href {https://doi.org/10.1103/PhysRevB.103.125131} {\bibfield  {journal} {\bibinfo  {journal} {Phys. Rev. B}\ }\textbf {\bibinfo {volume} {103}},\ \bibinfo {pages} {125131} (\bibinfo {year} {2021})}\BibitemShut {NoStop}%
\bibitem [{\citenamefont {Hosen}\ \emph {et~al.}(2018)\citenamefont {Hosen}, \citenamefont {Dhakal}, \citenamefont {Dimitri}, \citenamefont {Maldonado}, \citenamefont {Aperis}, \citenamefont {Kabir}, \citenamefont {Sims}, \citenamefont {Riseborough}, \citenamefont {Oppeneer}, \citenamefont {Kaczorowski}, \citenamefont {Durakiewicz},\ and\ \citenamefont {Neupane}}]{hosenDiscoveryTopologicalNodalline2018}%
  \BibitemOpen
  \bibfield  {author} {\bibinfo {author} {\bibfnamefont {M.~M.}\ \bibnamefont {Hosen}}, \bibinfo {author} {\bibfnamefont {G.}~\bibnamefont {Dhakal}}, \bibinfo {author} {\bibfnamefont {K.}~\bibnamefont {Dimitri}}, \bibinfo {author} {\bibfnamefont {P.}~\bibnamefont {Maldonado}}, \bibinfo {author} {\bibfnamefont {A.}~\bibnamefont {Aperis}}, \bibinfo {author} {\bibfnamefont {F.}~\bibnamefont {Kabir}}, \bibinfo {author} {\bibfnamefont {C.}~\bibnamefont {Sims}}, \bibinfo {author} {\bibfnamefont {P.}~\bibnamefont {Riseborough}}, \bibinfo {author} {\bibfnamefont {P.~M.}\ \bibnamefont {Oppeneer}}, \bibinfo {author} {\bibfnamefont {D.}~\bibnamefont {Kaczorowski}}, \bibinfo {author} {\bibfnamefont {T.}~\bibnamefont {Durakiewicz}},\ and\ \bibinfo {author} {\bibfnamefont {M.}~\bibnamefont {Neupane}},\ }\href {https://doi.org/10.1038/s41598-018-31296-7} {\bibfield  {journal} {\bibinfo  {journal} {Sci. Rep.}\ }\textbf {\bibinfo {volume} {8}},\ \bibinfo {pages} {13283} (\bibinfo {year} {2018})}\BibitemShut {NoStop}%
\bibitem [{\citenamefont {Ikeda}\ \emph {et~al.}(2022)\citenamefont {Ikeda}, \citenamefont {Saha}, \citenamefont {Graf}, \citenamefont {Saraf}, \citenamefont {Sokratov}, \citenamefont {Hu}, \citenamefont {Takahashi}, \citenamefont {Yamane}, \citenamefont {Jayaraj}, \citenamefont {S{\l}awi\'{n}ska}, \citenamefont {Nardelli}, \citenamefont {Yonezawa}, \citenamefont {Maeno},\ and\ \citenamefont {Paglione}}]{ikedaQuasitwodimensionalFermiSurface2022}%
  \BibitemOpen
  \bibfield  {author} {\bibinfo {author} {\bibfnamefont {A.}~\bibnamefont {Ikeda}}, \bibinfo {author} {\bibfnamefont {S.~R.}\ \bibnamefont {Saha}}, \bibinfo {author} {\bibfnamefont {D.}~\bibnamefont {Graf}}, \bibinfo {author} {\bibfnamefont {P.}~\bibnamefont {Saraf}}, \bibinfo {author} {\bibfnamefont {D.~S.}\ \bibnamefont {Sokratov}}, \bibinfo {author} {\bibfnamefont {Y.}~\bibnamefont {Hu}}, \bibinfo {author} {\bibfnamefont {H.}~\bibnamefont {Takahashi}}, \bibinfo {author} {\bibfnamefont {S.}~\bibnamefont {Yamane}}, \bibinfo {author} {\bibfnamefont {A.}~\bibnamefont {Jayaraj}}, \bibinfo {author} {\bibfnamefont {J.}~\bibnamefont {S{\l}awi\'{n}ska}}, \bibinfo {author} {\bibfnamefont {M.~B.}\ \bibnamefont {Nardelli}}, \bibinfo {author} {\bibfnamefont {S.}~\bibnamefont {Yonezawa}}, \bibinfo {author} {\bibfnamefont {Y.}~\bibnamefont {Maeno}},\ and\ \bibinfo {author} {\bibfnamefont {J.}~\bibnamefont {Paglione}},\ }\href {https://doi.org/10.1103/PhysRevB.106.075151} {\bibfield  {journal} {\bibinfo  {journal} {Phys. Rev. B}\ }\textbf {\bibinfo {volume} {106}},\ \bibinfo {pages} {075151} (\bibinfo {year} {2022})}\BibitemShut {NoStop}%
\bibitem [{\citenamefont {Chuang}\ \emph {et~al.}(2022)\citenamefont {Chuang}, \citenamefont {Souma}, \citenamefont {Moriya}, \citenamefont {Nakayama}, \citenamefont {Ikeda}, \citenamefont {Kawaguchi}, \citenamefont {Obata}, \citenamefont {Saha}, \citenamefont {Takahashi}, \citenamefont {Kitagawa}, \citenamefont {Ishida}, \citenamefont {Tanaka}, \citenamefont {Kitamura}, \citenamefont {Horiba}, \citenamefont {Kumigashira}, \citenamefont {Takahashi}, \citenamefont {Yonezawa}, \citenamefont {Paglione}, \citenamefont {Maeno},\ and\ \citenamefont {Sato}}]{chuangFermiologyTopologicalLinenodal2022}%
  \BibitemOpen
  \bibfield  {author} {\bibinfo {author} {\bibfnamefont {C.-W.}\ \bibnamefont {Chuang}}, \bibinfo {author} {\bibfnamefont {S.}~\bibnamefont {Souma}}, \bibinfo {author} {\bibfnamefont {A.}~\bibnamefont {Moriya}}, \bibinfo {author} {\bibfnamefont {K.}~\bibnamefont {Nakayama}}, \bibinfo {author} {\bibfnamefont {A.}~\bibnamefont {Ikeda}}, \bibinfo {author} {\bibfnamefont {M.}~\bibnamefont {Kawaguchi}}, \bibinfo {author} {\bibfnamefont {K.}~\bibnamefont {Obata}}, \bibinfo {author} {\bibfnamefont {S.~R.}\ \bibnamefont {Saha}}, \bibinfo {author} {\bibfnamefont {H.}~\bibnamefont {Takahashi}}, \bibinfo {author} {\bibfnamefont {S.}~\bibnamefont {Kitagawa}}, \bibinfo {author} {\bibfnamefont {K.}~\bibnamefont {Ishida}}, \bibinfo {author} {\bibfnamefont {K.}~\bibnamefont {Tanaka}}, \bibinfo {author} {\bibfnamefont {M.}~\bibnamefont {Kitamura}}, \bibinfo {author} {\bibfnamefont {K.}~\bibnamefont {Horiba}}, \bibinfo {author} {\bibfnamefont {H.}~\bibnamefont {Kumigashira}}, \bibinfo {author} {\bibfnamefont {T.}~\bibnamefont {Takahashi}}, \bibinfo {author} {\bibfnamefont {S.}~\bibnamefont {Yonezawa}}, \bibinfo {author} {\bibfnamefont {J.}~\bibnamefont {Paglione}}, \bibinfo {author} {\bibfnamefont {Y.}~\bibnamefont {Maeno}},\ and\ \bibinfo {author} {\bibfnamefont {T.}~\bibnamefont {Sato}},\ }\href {https://doi.org/10.1103/PhysRevMaterials.6.104203} {\bibfield  {journal} {\bibinfo  {journal} {Phys. Rev. Mater.}\ }\textbf {\bibinfo {volume} {6}},\ \bibinfo {pages} {104203} (\bibinfo {year} {2022})}\BibitemShut {NoStop}%
\bibitem [{\citenamefont {Ikeda}\ \emph {et~al.}(2020)\citenamefont {Ikeda}, \citenamefont {Kawaguchi}, \citenamefont {Koibuchi}, \citenamefont {Hashimoto}, \citenamefont {Kawakami}, \citenamefont {Yonezawa}, \citenamefont {Sato},\ and\ \citenamefont {Maeno}}]{ikedaSuperconductivityNonsymmorphicLinenodal2020}%
  \BibitemOpen
  \bibfield  {author} {\bibinfo {author} {\bibfnamefont {A.}~\bibnamefont {Ikeda}}, \bibinfo {author} {\bibfnamefont {M.}~\bibnamefont {Kawaguchi}}, \bibinfo {author} {\bibfnamefont {S.}~\bibnamefont {Koibuchi}}, \bibinfo {author} {\bibfnamefont {T.}~\bibnamefont {Hashimoto}}, \bibinfo {author} {\bibfnamefont {T.}~\bibnamefont {Kawakami}}, \bibinfo {author} {\bibfnamefont {S.}~\bibnamefont {Yonezawa}}, \bibinfo {author} {\bibfnamefont {M.}~\bibnamefont {Sato}},\ and\ \bibinfo {author} {\bibfnamefont {Y.}~\bibnamefont {Maeno}},\ }\href {https://doi.org/10.1103/PhysRevMaterials.4.041801} {\bibfield  {journal} {\bibinfo  {journal} {Phys. Rev. Mater.}\ }\textbf {\bibinfo {volume} {4}},\ \bibinfo {pages} {041801(R)} (\bibinfo {year} {2020})}\BibitemShut {NoStop}%
\bibitem [{\citenamefont {Takahashi}\ \emph {et~al.}(2021)\citenamefont {Takahashi}, \citenamefont {Kitagawa}, \citenamefont {Ishida}, \citenamefont {Kawaguchi}, \citenamefont {Ikeda}, \citenamefont {Yonezawa},\ and\ \citenamefont {Maeno}}]{takahashiSWaveSuperconductivityDirac2021}%
  \BibitemOpen
  \bibfield  {author} {\bibinfo {author} {\bibfnamefont {H.}~\bibnamefont {Takahashi}}, \bibinfo {author} {\bibfnamefont {S.}~\bibnamefont {Kitagawa}}, \bibinfo {author} {\bibfnamefont {K.}~\bibnamefont {Ishida}}, \bibinfo {author} {\bibfnamefont {M.}~\bibnamefont {Kawaguchi}}, \bibinfo {author} {\bibfnamefont {A.}~\bibnamefont {Ikeda}}, \bibinfo {author} {\bibfnamefont {S.}~\bibnamefont {Yonezawa}},\ and\ \bibinfo {author} {\bibfnamefont {Y.}~\bibnamefont {Maeno}},\ }\href {https://doi.org/10.7566/JPSJ.90.073702} {\bibfield  {journal} {\bibinfo  {journal} {J. Phys. Soc. Jpn.}\ }\textbf {\bibinfo {volume} {90}},\ \bibinfo {pages} {073702} (\bibinfo {year} {2021})}\BibitemShut {NoStop}%
\bibitem [{\citenamefont {Duan}\ \emph {et~al.}(2022)\citenamefont {Duan}, \citenamefont {Zhang}, \citenamefont {Kumar}, \citenamefont {Su}, \citenamefont {Zhou}, \citenamefont {Nie}, \citenamefont {Chen}, \citenamefont {Smidman}, \citenamefont {Cao}, \citenamefont {Song},\ and\ \citenamefont {Yuan}}]{PhysRevB.106.214521}%
  \BibitemOpen
  \bibfield  {author} {\bibinfo {author} {\bibfnamefont {W.}~\bibnamefont {Duan}}, \bibinfo {author} {\bibfnamefont {J.}~\bibnamefont {Zhang}}, \bibinfo {author} {\bibfnamefont {R.}~\bibnamefont {Kumar}}, \bibinfo {author} {\bibfnamefont {H.}~\bibnamefont {Su}}, \bibinfo {author} {\bibfnamefont {Y.}~\bibnamefont {Zhou}}, \bibinfo {author} {\bibfnamefont {Z.}~\bibnamefont {Nie}}, \bibinfo {author} {\bibfnamefont {Y.}~\bibnamefont {Chen}}, \bibinfo {author} {\bibfnamefont {M.}~\bibnamefont {Smidman}}, \bibinfo {author} {\bibfnamefont {C.}~\bibnamefont {Cao}}, \bibinfo {author} {\bibfnamefont {Y.}~\bibnamefont {Song}},\ and\ \bibinfo {author} {\bibfnamefont {H.}~\bibnamefont {Yuan}},\ }\href {https://doi.org/10.1103/PhysRevB.106.214521} {\bibfield  {journal} {\bibinfo  {journal} {Phys. Rev. B}\ }\textbf {\bibinfo {volume} {106}},\ \bibinfo {pages} {214521} (\bibinfo {year} {2022})}\BibitemShut {NoStop}%
\bibitem [{\citenamefont {Ikeda}\ \emph {et~al.}(2024)\citenamefont {Ikeda}, \citenamefont {Yonezawa},\ and\ \citenamefont {Maeno}}]{ikedaSuppressionSuperconductingTransition2024}%
  \BibitemOpen
  \bibfield  {author} {\bibinfo {author} {\bibfnamefont {A.}~\bibnamefont {Ikeda}}, \bibinfo {author} {\bibfnamefont {S.}~\bibnamefont {Yonezawa}},\ and\ \bibinfo {author} {\bibfnamefont {Y.}~\bibnamefont {Maeno}},\ }\href {https://doi.org/10.7566/JPSJ.93.075002} {\bibfield  {journal} {\bibinfo  {journal} {J. Phys. Soc. Jpn.}\ }\textbf {\bibinfo {volume} {93}},\ \bibinfo {pages} {075002} (\bibinfo {year} {2024})}\BibitemShut {NoStop}%
\bibitem [{\citenamefont {Kitagawa}\ \emph {et~al.}(2021)\citenamefont {Kitagawa}, \citenamefont {Ishida}, \citenamefont {Ikeda}, \citenamefont {Kawaguchi}, \citenamefont {Yonezawa},\ and\ \citenamefont {Maeno}}]{kitagawaPeakSuperconductingTransition2021}%
  \BibitemOpen
  \bibfield  {author} {\bibinfo {author} {\bibfnamefont {S.}~\bibnamefont {Kitagawa}}, \bibinfo {author} {\bibfnamefont {K.}~\bibnamefont {Ishida}}, \bibinfo {author} {\bibfnamefont {A.}~\bibnamefont {Ikeda}}, \bibinfo {author} {\bibfnamefont {M.}~\bibnamefont {Kawaguchi}}, \bibinfo {author} {\bibfnamefont {S.}~\bibnamefont {Yonezawa}},\ and\ \bibinfo {author} {\bibfnamefont {Y.}~\bibnamefont {Maeno}},\ }\href {https://doi.org/10.1103/PhysRevB.104.L060504} {\bibfield  {journal} {\bibinfo  {journal} {Phys. Rev. B}\ }\textbf {\bibinfo {volume} {104}},\ \bibinfo {pages} {L060504} (\bibinfo {year} {2021})}\BibitemShut {NoStop}%
\bibitem [{\citenamefont {Takahashi}\ \emph {et~al.}(2024)\citenamefont {Takahashi}, \citenamefont {Kitagawa}, \citenamefont {Ishida}, \citenamefont {Ikeda}, \citenamefont {Saha}, \citenamefont {Yonezawa}, \citenamefont {Paglione},\ and\ \citenamefont {Maeno}}]{PhysRevB.109.L100501}%
  \BibitemOpen
  \bibfield  {author} {\bibinfo {author} {\bibfnamefont {H.}~\bibnamefont {Takahashi}}, \bibinfo {author} {\bibfnamefont {S.}~\bibnamefont {Kitagawa}}, \bibinfo {author} {\bibfnamefont {K.}~\bibnamefont {Ishida}}, \bibinfo {author} {\bibfnamefont {A.}~\bibnamefont {Ikeda}}, \bibinfo {author} {\bibfnamefont {S.~R.}\ \bibnamefont {Saha}}, \bibinfo {author} {\bibfnamefont {S.}~\bibnamefont {Yonezawa}}, \bibinfo {author} {\bibfnamefont {J.}~\bibnamefont {Paglione}},\ and\ \bibinfo {author} {\bibfnamefont {Y.}~\bibnamefont {Maeno}},\ }\href {https://doi.org/10.1103/PhysRevB.109.L100501} {\bibfield  {journal} {\bibinfo  {journal} {Phys. Rev. B}\ }\textbf {\bibinfo {volume} {109}},\ \bibinfo {pages} {L100501} (\bibinfo {year} {2024})}\BibitemShut {NoStop}%
\bibitem [{\citenamefont {Momma}\ and\ \citenamefont {Izumi}(2011)}]{Momma:db5098}%
  \BibitemOpen
  \bibfield  {author} {\bibinfo {author} {\bibfnamefont {K.}~\bibnamefont {Momma}}\ and\ \bibinfo {author} {\bibfnamefont {F.}~\bibnamefont {Izumi}},\ }\href {https://doi.org/10.1107/S0021889811038970} {\bibfield  {journal} {\bibinfo  {journal} {J. Appl. Crystallogr.}\ }\textbf {\bibinfo {volume} {44}},\ \bibinfo {pages} {1272} (\bibinfo {year} {2011})}\BibitemShut {NoStop}%
\bibitem [{\citenamefont {Hirao}\ \emph {et~al.}(2020)\citenamefont {Hirao}, \citenamefont {Kawaguchi}, \citenamefont {Hirose}, \citenamefont {Shimizu}, \citenamefont {Ohtani},\ and\ \citenamefont {Ohishi}}]{hiraoNewDevelopmentsHighpressure2020}%
  \BibitemOpen
  \bibfield  {author} {\bibinfo {author} {\bibfnamefont {N.}~\bibnamefont {Hirao}}, \bibinfo {author} {\bibfnamefont {S.~I.}\ \bibnamefont {Kawaguchi}}, \bibinfo {author} {\bibfnamefont {K.}~\bibnamefont {Hirose}}, \bibinfo {author} {\bibfnamefont {K.}~\bibnamefont {Shimizu}}, \bibinfo {author} {\bibfnamefont {E.}~\bibnamefont {Ohtani}},\ and\ \bibinfo {author} {\bibfnamefont {Y.}~\bibnamefont {Ohishi}},\ }\href {https://doi.org/10.1063/1.5126038} {\bibfield  {journal} {\bibinfo  {journal} {MRE}\ }\textbf {\bibinfo {volume} {5}},\ \bibinfo {pages} {018403} (\bibinfo {year} {2020})}\BibitemShut {NoStop}%
\bibitem [{\citenamefont {Mao}\ \emph {et~al.}(1986)\citenamefont {Mao}, \citenamefont {Xu},\ and\ \citenamefont {Bell}}]{https://doi.org/10.1029/JB091iB05p04673}%
  \BibitemOpen
  \bibfield  {author} {\bibinfo {author} {\bibfnamefont {H.~K.}\ \bibnamefont {Mao}}, \bibinfo {author} {\bibfnamefont {J.}~\bibnamefont {Xu}},\ and\ \bibinfo {author} {\bibfnamefont {P.~M.}\ \bibnamefont {Bell}},\ }\href {https://doi.org/10.1029/JB091iB05p04673} {\bibfield  {journal} {\bibinfo  {journal} {J. Geophys. Res.: Solid Earth}\ }\textbf {\bibinfo {volume} {91}},\ \bibinfo {pages} {4673} (\bibinfo {year} {1986})}\BibitemShut {NoStop}%
\bibitem [{\citenamefont {SETO}\ \emph {et~al.}(2010)\citenamefont {SETO}, \citenamefont {{NISHIO-HAMANE}}, \citenamefont {NAGAI},\ and\ \citenamefont {SATA}}]{setoDevelopmentSoftwareSuite2010}%
  \BibitemOpen
  \bibfield  {author} {\bibinfo {author} {\bibfnamefont {Y.}~\bibnamefont {SETO}}, \bibinfo {author} {\bibfnamefont {D.}~\bibnamefont {{NISHIO-HAMANE}}}, \bibinfo {author} {\bibfnamefont {T.}~\bibnamefont {NAGAI}},\ and\ \bibinfo {author} {\bibfnamefont {N.}~\bibnamefont {SATA}},\ }\href {https://doi.org/10.4131/jshpreview.20.269} {\bibfield  {journal} {\bibinfo  {journal} {Rev. High Press. Sci. Technol.}\ }\textbf {\bibinfo {volume} {20}},\ \bibinfo {pages} {269} (\bibinfo {year} {2010})}\BibitemShut {NoStop}%
\bibitem [{\citenamefont {Rietveld}(1969)}]{Rietveld:a07067}%
  \BibitemOpen
  \bibfield  {author} {\bibinfo {author} {\bibfnamefont {H.~M.}\ \bibnamefont {Rietveld}},\ }\href {https://doi.org/10.1107/S0021889869006558} {\bibfield  {journal} {\bibinfo  {journal} {J. Appl. Crystallogr.}\ }\textbf {\bibinfo {volume} {2}},\ \bibinfo {pages} {65} (\bibinfo {year} {1969})}\BibitemShut {NoStop}%
\bibitem [{\citenamefont {Pawley}(1981)}]{Pawley:a20546}%
  \BibitemOpen
  \bibfield  {author} {\bibinfo {author} {\bibfnamefont {G.~S.}\ \bibnamefont {Pawley}},\ }\href {https://doi.org/10.1107/S0021889881009618} {\bibfield  {journal} {\bibinfo  {journal} {J. Appl. Crystallogr.}\ }\textbf {\bibinfo {volume} {14}},\ \bibinfo {pages} {357} (\bibinfo {year} {1981})}\BibitemShut {NoStop}%
\bibitem [{\citenamefont {Giannozzi}\ \emph {et~al.}(2009)\citenamefont {Giannozzi}, \citenamefont {Baroni}, \citenamefont {Bonini}, \citenamefont {Calandra}, \citenamefont {Car}, \citenamefont {Cavazzoni}, \citenamefont {Ceresoli}, \citenamefont {Chiarotti}, \citenamefont {Cococcioni}, \citenamefont {Dabo}, \citenamefont {Dal~Corso}, \citenamefont {{de Gironcoli}}, \citenamefont {Fabris}, \citenamefont {Fratesi}, \citenamefont {Gebauer}, \citenamefont {Gerstmann}, \citenamefont {Gougoussis}, \citenamefont {Kokalj}, \citenamefont {Lazzeri}, \citenamefont {{Martin-Samos}}, \citenamefont {Marzari}, \citenamefont {Mauri}, \citenamefont {Mazzarello}, \citenamefont {Paolini}, \citenamefont {Pasquarello}, \citenamefont {Paulatto}, \citenamefont {Sbraccia}, \citenamefont {Scandolo}, \citenamefont {Sclauzero}, \citenamefont {Seitsonen}, \citenamefont {Smogunov}, \citenamefont {Umari},\ and\ \citenamefont {Wentzcovitch}}]{giannozziQUANTUMESPRESSOModular2009}%
  \BibitemOpen
  \bibfield  {author} {\bibinfo {author} {\bibfnamefont {P.}~\bibnamefont {Giannozzi}}, \bibinfo {author} {\bibfnamefont {S.}~\bibnamefont {Baroni}}, \bibinfo {author} {\bibfnamefont {N.}~\bibnamefont {Bonini}}, \bibinfo {author} {\bibfnamefont {M.}~\bibnamefont {Calandra}}, \bibinfo {author} {\bibfnamefont {R.}~\bibnamefont {Car}}, \bibinfo {author} {\bibfnamefont {C.}~\bibnamefont {Cavazzoni}}, \bibinfo {author} {\bibfnamefont {D.}~\bibnamefont {Ceresoli}}, \bibinfo {author} {\bibfnamefont {G.~L.}\ \bibnamefont {Chiarotti}}, \bibinfo {author} {\bibfnamefont {M.}~\bibnamefont {Cococcioni}}, \bibinfo {author} {\bibfnamefont {I.}~\bibnamefont {Dabo}}, \bibinfo {author} {\bibfnamefont {A.}~\bibnamefont {Dal~Corso}}, \bibinfo {author} {\bibfnamefont {S.}~\bibnamefont {{de Gironcoli}}}, \bibinfo {author} {\bibfnamefont {S.}~\bibnamefont {Fabris}}, \bibinfo {author} {\bibfnamefont {G.}~\bibnamefont {Fratesi}}, \bibinfo {author} {\bibfnamefont {R.}~\bibnamefont {Gebauer}}, \bibinfo {author} {\bibfnamefont {U.}~\bibnamefont {Gerstmann}}, \bibinfo {author} {\bibfnamefont {C.}~\bibnamefont {Gougoussis}}, \bibinfo {author} {\bibfnamefont {A.}~\bibnamefont {Kokalj}}, \bibinfo {author} {\bibfnamefont {M.}~\bibnamefont {Lazzeri}}, \bibinfo {author} {\bibfnamefont {L.}~\bibnamefont {{Martin-Samos}}}, \bibinfo {author} {\bibfnamefont {N.}~\bibnamefont {Marzari}}, \bibinfo {author} {\bibfnamefont {F.}~\bibnamefont {Mauri}}, \bibinfo {author} {\bibfnamefont {R.}~\bibnamefont {Mazzarello}}, \bibinfo {author} {\bibfnamefont {S.}~\bibnamefont {Paolini}}, \bibinfo {author} {\bibfnamefont {A.}~\bibnamefont {Pasquarello}}, \bibinfo {author} {\bibfnamefont {L.}~\bibnamefont {Paulatto}}, \bibinfo {author} {\bibfnamefont {C.}~\bibnamefont {Sbraccia}}, \bibinfo {author} {\bibfnamefont {S.}~\bibnamefont {Scandolo}}, \bibinfo {author} {\bibfnamefont {G.}~\bibnamefont {Sclauzero}}, \bibinfo {author} {\bibfnamefont {A.~P.}\ \bibnamefont {Seitsonen}}, \bibinfo {author} {\bibfnamefont {A.}~\bibnamefont {Smogunov}}, \bibinfo {author} {\bibfnamefont {P.}~\bibnamefont {Umari}},\ and\ \bibinfo {author} {\bibfnamefont {R.~M.}\ \bibnamefont {Wentzcovitch}},\ }\href {https://doi.org/10.1088/0953-8984/21/39/395502} {\bibfield  {journal} {\bibinfo  {journal} {J. Phys.: Condens. Matter}\ }\textbf {\bibinfo {volume} {21}},\ \bibinfo {pages} {395502} (\bibinfo {year} {2009})}\BibitemShut {NoStop}%
\bibitem [{\citenamefont {Giannozzi}\ \emph {et~al.}(2017)\citenamefont {Giannozzi}, \citenamefont {Andreussi}, \citenamefont {Brumme}, \citenamefont {Bunau}, \citenamefont {Buongiorno~Nardelli}, \citenamefont {Calandra}, \citenamefont {Car}, \citenamefont {Cavazzoni}, \citenamefont {Ceresoli}, \citenamefont {Cococcioni}, \citenamefont {Colonna}, \citenamefont {Carnimeo}, \citenamefont {Dal~Corso}, \citenamefont {{de Gironcoli}}, \citenamefont {Delugas}, \citenamefont {DiStasio}, \citenamefont {Ferretti}, \citenamefont {Floris}, \citenamefont {Fratesi}, \citenamefont {Fugallo}, \citenamefont {Gebauer}, \citenamefont {Gerstmann}, \citenamefont {Giustino}, \citenamefont {Gorni}, \citenamefont {Jia}, \citenamefont {Kawamura}, \citenamefont {Ko}, \citenamefont {Kokalj}, \citenamefont {K{\"u}{\c c}{\"u}kbenli}, \citenamefont {Lazzeri}, \citenamefont {Marsili}, \citenamefont {Marzari}, \citenamefont {Mauri}, \citenamefont {Nguyen}, \citenamefont {Nguyen}, \citenamefont {{Otero-de-la-Roza}}, \citenamefont {Paulatto}, \citenamefont {Ponc{\'e}}, \citenamefont {Rocca}, \citenamefont {Sabatini}, \citenamefont {Santra}, \citenamefont {Schlipf}, \citenamefont {Seitsonen}, \citenamefont {Smogunov}, \citenamefont {Timrov}, \citenamefont {Thonhauser}, \citenamefont {Umari}, \citenamefont {Vast}, \citenamefont {Wu},\ and\ \citenamefont {Baroni}}]{giannozziAdvancedCapabilitiesMaterials2017}%
  \BibitemOpen
  \bibfield  {author} {\bibinfo {author} {\bibfnamefont {P.}~\bibnamefont {Giannozzi}}, \bibinfo {author} {\bibfnamefont {O.}~\bibnamefont {Andreussi}}, \bibinfo {author} {\bibfnamefont {T.}~\bibnamefont {Brumme}}, \bibinfo {author} {\bibfnamefont {O.}~\bibnamefont {Bunau}}, \bibinfo {author} {\bibfnamefont {M.}~\bibnamefont {Buongiorno~Nardelli}}, \bibinfo {author} {\bibfnamefont {M.}~\bibnamefont {Calandra}}, \bibinfo {author} {\bibfnamefont {R.}~\bibnamefont {Car}}, \bibinfo {author} {\bibfnamefont {C.}~\bibnamefont {Cavazzoni}}, \bibinfo {author} {\bibfnamefont {D.}~\bibnamefont {Ceresoli}}, \bibinfo {author} {\bibfnamefont {M.}~\bibnamefont {Cococcioni}}, \bibinfo {author} {\bibfnamefont {N.}~\bibnamefont {Colonna}}, \bibinfo {author} {\bibfnamefont {I.}~\bibnamefont {Carnimeo}}, \bibinfo {author} {\bibfnamefont {A.}~\bibnamefont {Dal~Corso}}, \bibinfo {author} {\bibfnamefont {S.}~\bibnamefont {{de Gironcoli}}}, \bibinfo {author} {\bibfnamefont {P.}~\bibnamefont {Delugas}}, \bibinfo {author} {\bibfnamefont {R.~A.~J.}\ \bibnamefont {DiStasio}}, \bibinfo {author} {\bibfnamefont {A.}~\bibnamefont {Ferretti}}, \bibinfo {author} {\bibfnamefont {A.}~\bibnamefont {Floris}}, \bibinfo {author} {\bibfnamefont {G.}~\bibnamefont {Fratesi}}, \bibinfo {author} {\bibfnamefont {G.}~\bibnamefont {Fugallo}}, \bibinfo {author} {\bibfnamefont {R.}~\bibnamefont {Gebauer}}, \bibinfo {author} {\bibfnamefont {U.}~\bibnamefont {Gerstmann}}, \bibinfo {author} {\bibfnamefont {F.}~\bibnamefont {Giustino}}, \bibinfo {author} {\bibfnamefont {T.}~\bibnamefont {Gorni}}, \bibinfo {author} {\bibfnamefont {J.}~\bibnamefont {Jia}}, \bibinfo {author} {\bibfnamefont {M.}~\bibnamefont {Kawamura}}, \bibinfo {author} {\bibfnamefont {H.-Y.}\ \bibnamefont {Ko}}, \bibinfo {author} {\bibfnamefont {A.}~\bibnamefont {Kokalj}}, \bibinfo {author} {\bibfnamefont {E.}~\bibnamefont {K{\"u}{\c c}{\"u}kbenli}}, \bibinfo {author} {\bibfnamefont {M.}~\bibnamefont {Lazzeri}}, \bibinfo {author} {\bibfnamefont {M.}~\bibnamefont {Marsili}}, \bibinfo {author} {\bibfnamefont {N.}~\bibnamefont {Marzari}}, \bibinfo {author} {\bibfnamefont {F.}~\bibnamefont {Mauri}}, \bibinfo {author} {\bibfnamefont {N.~L.}\ \bibnamefont {Nguyen}}, \bibinfo {author} {\bibfnamefont {H.-V.}\ \bibnamefont {Nguyen}}, \bibinfo {author} {\bibfnamefont {A.}~\bibnamefont {{Otero-de-la-Roza}}}, \bibinfo {author} {\bibfnamefont {L.}~\bibnamefont {Paulatto}}, \bibinfo {author} {\bibfnamefont {S.}~\bibnamefont {Ponc{\'e}}}, \bibinfo {author} {\bibfnamefont {D.}~\bibnamefont {Rocca}}, \bibinfo {author} {\bibfnamefont {R.}~\bibnamefont {Sabatini}}, \bibinfo {author} {\bibfnamefont {B.}~\bibnamefont {Santra}}, \bibinfo {author} {\bibfnamefont {M.}~\bibnamefont {Schlipf}}, \bibinfo {author} {\bibfnamefont {A.~P.}\ \bibnamefont {Seitsonen}}, \bibinfo {author} {\bibfnamefont {A.}~\bibnamefont {Smogunov}}, \bibinfo {author} {\bibfnamefont {I.}~\bibnamefont {Timrov}}, \bibinfo {author} {\bibfnamefont {T.}~\bibnamefont {Thonhauser}}, \bibinfo {author} {\bibfnamefont {P.}~\bibnamefont {Umari}}, \bibinfo {author} {\bibfnamefont {N.}~\bibnamefont {Vast}}, \bibinfo {author} {\bibfnamefont {X.}~\bibnamefont {Wu}},\ and\ \bibinfo {author} {\bibfnamefont {S.}~\bibnamefont {Baroni}},\ }\href {https://doi.org/10.1088/1361-648X/aa8f79} {\bibfield  {journal} {\bibinfo  {journal} {J. Phys.: Condens. Matter}\ }\textbf {\bibinfo {volume} {29}},\ \bibinfo {pages} {465901} (\bibinfo {year} {2017})}\BibitemShut {NoStop}%
\bibitem [{\citenamefont {Perdew}\ \emph {et~al.}(1996)\citenamefont {Perdew}, \citenamefont {Burke},\ and\ \citenamefont {Ernzerhof}}]{PhysRevLett.77.3865}%
  \BibitemOpen
  \bibfield  {author} {\bibinfo {author} {\bibfnamefont {J.~P.}\ \bibnamefont {Perdew}}, \bibinfo {author} {\bibfnamefont {K.}~\bibnamefont {Burke}},\ and\ \bibinfo {author} {\bibfnamefont {M.}~\bibnamefont {Ernzerhof}},\ }\href {https://doi.org/10.1103/PhysRevLett.77.3865} {\bibfield  {journal} {\bibinfo  {journal} {Phys. Rev. Lett.}\ }\textbf {\bibinfo {volume} {77}},\ \bibinfo {pages} {3865} (\bibinfo {year} {1996})}\BibitemShut {NoStop}%
\bibitem [{\citenamefont {Dal~Corso}(2014)}]{dalcorsoPseudopotentialsPeriodicTable2014}%
  \BibitemOpen
  \bibfield  {author} {\bibinfo {author} {\bibfnamefont {A.}~\bibnamefont {Dal~Corso}},\ }\href {https://doi.org/10.1016/j.commatsci.2014.07.043} {\bibfield  {journal} {\bibinfo  {journal} {Comput. Mater. Sci.}\ }\textbf {\bibinfo {volume} {95}},\ \bibinfo {pages} {337} (\bibinfo {year} {2014})}\BibitemShut {NoStop}%
\bibitem [{\citenamefont {Mostofi}\ \emph {et~al.}(2014)\citenamefont {Mostofi}, \citenamefont {Yates}, \citenamefont {Pizzi}, \citenamefont {Lee}, \citenamefont {Souza}, \citenamefont {Vanderbilt},\ and\ \citenamefont {Marzari}}]{mostofiUpdatedVersionWannier902014}%
  \BibitemOpen
  \bibfield  {author} {\bibinfo {author} {\bibfnamefont {A.~A.}\ \bibnamefont {Mostofi}}, \bibinfo {author} {\bibfnamefont {J.~R.}\ \bibnamefont {Yates}}, \bibinfo {author} {\bibfnamefont {G.}~\bibnamefont {Pizzi}}, \bibinfo {author} {\bibfnamefont {Y.-S.}\ \bibnamefont {Lee}}, \bibinfo {author} {\bibfnamefont {I.}~\bibnamefont {Souza}}, \bibinfo {author} {\bibfnamefont {D.}~\bibnamefont {Vanderbilt}},\ and\ \bibinfo {author} {\bibfnamefont {N.}~\bibnamefont {Marzari}},\ }\href {https://doi.org/10.1016/j.cpc.2014.05.003} {\bibfield  {journal} {\bibinfo  {journal} {Comput. Phys. Commun.}\ }\textbf {\bibinfo {volume} {185}},\ \bibinfo {pages} {2309} (\bibinfo {year} {2014})}\BibitemShut {NoStop}%
\bibitem [{\citenamefont {Wu}\ \emph {et~al.}(2018)\citenamefont {Wu}, \citenamefont {Zhang}, \citenamefont {Song}, \citenamefont {Troyer},\ and\ \citenamefont {Soluyanov}}]{wuWannierToolsOpensourceSoftware2018}%
  \BibitemOpen
  \bibfield  {author} {\bibinfo {author} {\bibfnamefont {Q.}~\bibnamefont {Wu}}, \bibinfo {author} {\bibfnamefont {S.}~\bibnamefont {Zhang}}, \bibinfo {author} {\bibfnamefont {H.-F.}\ \bibnamefont {Song}}, \bibinfo {author} {\bibfnamefont {M.}~\bibnamefont {Troyer}},\ and\ \bibinfo {author} {\bibfnamefont {A.~A.}\ \bibnamefont {Soluyanov}},\ }\href {https://doi.org/10.1016/j.cpc.2017.09.033} {\bibfield  {journal} {\bibinfo  {journal} {Comput. Phys. Commun.}\ }\textbf {\bibinfo {volume} {224}},\ \bibinfo {pages} {405} (\bibinfo {year} {2018})}\BibitemShut {NoStop}%
\bibitem [{\citenamefont {Birch}(1947)}]{PhysRev.71.809}%
  \BibitemOpen
  \bibfield  {author} {\bibinfo {author} {\bibfnamefont {F.}~\bibnamefont {Birch}},\ }\href {https://doi.org/10.1103/PhysRev.71.809} {\bibfield  {journal} {\bibinfo  {journal} {Phys. Rev.}\ }\textbf {\bibinfo {volume} {71}},\ \bibinfo {pages} {809} (\bibinfo {year} {1947})}\BibitemShut {NoStop}%
\bibitem [{\citenamefont {Christy}(1995)}]{Christy:ab0333}%
  \BibitemOpen
  \bibfield  {author} {\bibinfo {author} {\bibfnamefont {A.~G.}\ \bibnamefont {Christy}},\ }\href {https://doi.org/10.1107/S0108768195001728} {\bibfield  {journal} {\bibinfo  {journal} {Acta Cryst. B}\ }\textbf {\bibinfo {volume} {51}},\ \bibinfo {pages} {753} (\bibinfo {year} {1995})}\BibitemShut {NoStop}%
\bibitem [{\citenamefont {Kaluarachchi}\ \emph {et~al.}(2017)\citenamefont {Kaluarachchi}, \citenamefont {Taufour}, \citenamefont {Sapkota}, \citenamefont {Borisov}, \citenamefont {Kong}, \citenamefont {Meier}, \citenamefont {Kothapalli}, \citenamefont {Ueland}, \citenamefont {Kreyssig}, \citenamefont {Valent{\'{\i}}}, \citenamefont {McQueeney}, \citenamefont {Goldman}, \citenamefont {Bud'ko},\ and\ \citenamefont {Canfield}}]{PhysRevB.96.140501}%
  \BibitemOpen
  \bibfield  {author} {\bibinfo {author} {\bibfnamefont {U.~S.}\ \bibnamefont {Kaluarachchi}}, \bibinfo {author} {\bibfnamefont {V.}~\bibnamefont {Taufour}}, \bibinfo {author} {\bibfnamefont {A.}~\bibnamefont {Sapkota}}, \bibinfo {author} {\bibfnamefont {V.}~\bibnamefont {Borisov}}, \bibinfo {author} {\bibfnamefont {T.}~\bibnamefont {Kong}}, \bibinfo {author} {\bibfnamefont {W.~R.}\ \bibnamefont {Meier}}, \bibinfo {author} {\bibfnamefont {K.}~\bibnamefont {Kothapalli}}, \bibinfo {author} {\bibfnamefont {B.~G.}\ \bibnamefont {Ueland}}, \bibinfo {author} {\bibfnamefont {A.}~\bibnamefont {Kreyssig}}, \bibinfo {author} {\bibfnamefont {R.}~\bibnamefont {Valent{\'{\i}}}}, \bibinfo {author} {\bibfnamefont {R.~J.}\ \bibnamefont {McQueeney}}, \bibinfo {author} {\bibfnamefont {A.~I.}\ \bibnamefont {Goldman}}, \bibinfo {author} {\bibfnamefont {S.~L.}\ \bibnamefont {Bud'ko}},\ and\ \bibinfo {author} {\bibfnamefont {P.~C.}\ \bibnamefont {Canfield}},\ }\href {https://doi.org/10.1103/PhysRevB.96.140501} {\bibfield  {journal} {\bibinfo  {journal} {Phys. Rev. B}\ }\textbf {\bibinfo {volume} {96}},\ \bibinfo {pages} {140501} (\bibinfo {year} {2017})}\BibitemShut {NoStop}%
\bibitem [{\citenamefont {Lorenz}\ \emph {et~al.}(2003)\citenamefont {Lorenz}, \citenamefont {Cmaidalka}, \citenamefont {Meng},\ and\ \citenamefont {Chu}}]{PhysRevB.68.014512}%
  \BibitemOpen
  \bibfield  {author} {\bibinfo {author} {\bibfnamefont {B.}~\bibnamefont {Lorenz}}, \bibinfo {author} {\bibfnamefont {J.}~\bibnamefont {Cmaidalka}}, \bibinfo {author} {\bibfnamefont {R.~L.}\ \bibnamefont {Meng}},\ and\ \bibinfo {author} {\bibfnamefont {C.~W.}\ \bibnamefont {Chu}},\ }\href {https://doi.org/10.1103/PhysRevB.68.014512} {\bibfield  {journal} {\bibinfo  {journal} {Phys. Rev. B}\ }\textbf {\bibinfo {volume} {68}},\ \bibinfo {pages} {014512} (\bibinfo {year} {2003})}\BibitemShut {NoStop}%
\bibitem [{\citenamefont {Walicka}\ \emph {et~al.}(2021)\citenamefont {Walicka}, \citenamefont {Guguchia}, \citenamefont {Lago}, \citenamefont {Blacque}, \citenamefont {Ma}, \citenamefont {Liu}, \citenamefont {Khasanov},\ and\ \citenamefont {{von Rohr}}}]{PhysRevResearch.3.033192}%
  \BibitemOpen
  \bibfield  {author} {\bibinfo {author} {\bibfnamefont {D.~I.}\ \bibnamefont {Walicka}}, \bibinfo {author} {\bibfnamefont {Z.}~\bibnamefont {Guguchia}}, \bibinfo {author} {\bibfnamefont {J.}~\bibnamefont {Lago}}, \bibinfo {author} {\bibfnamefont {O.}~\bibnamefont {Blacque}}, \bibinfo {author} {\bibfnamefont {K.}~\bibnamefont {Ma}}, \bibinfo {author} {\bibfnamefont {H.}~\bibnamefont {Liu}}, \bibinfo {author} {\bibfnamefont {R.}~\bibnamefont {Khasanov}},\ and\ \bibinfo {author} {\bibfnamefont {F.~O.}\ \bibnamefont {{von Rohr}}},\ }\href {https://doi.org/10.1103/PhysRevResearch.3.033192} {\bibfield  {journal} {\bibinfo  {journal} {Phys. Rev. Res.}\ }\textbf {\bibinfo {volume} {3}},\ \bibinfo {pages} {033192} (\bibinfo {year} {2021})}\BibitemShut {NoStop}%
\end{thebibliography}%

\end{document}